\documentclass[fleqn,usenatbib]{mnras}
\usepackage{verbatim}
\usepackage{hyperref}
\hypersetup{
    colorlinks=true,
    linkcolor=blue,
    filecolor=magenta,      
    urlcolor=cyan,
    citecolor=blue,
}

\usepackage{etoolbox,xparse}
\usepackage[T2A,T1]{fontenc}
\usepackage[utf8]{inputenc}
\usepackage[bulgarian,british]{babel}
\usepackage{datetime2}
\usepackage[usenames,dvipsnames]{xcolor}
\usepackage{rotating}

\usepackage{multirow}
\usepackage{booktabs}  
\usepackage{graphicx}
\usepackage{amsmath}
\usepackage{newtxtext}
\usepackage{newtxmath}
\usepackage[style=english,english=british]{csquotes}  
\usepackage{siunitx}  
\DeclareSIPostPower{\nominal}{N}
\DeclareSIQualifier{\earth}{\ensuremath{\oplus}}
\DeclareSIQualifier{\jupiter}{J}
\DeclareSIQualifier{\planet}{p}
\DeclareSIQualifier{\etoile}{\ensuremath{\star}}
\DeclareSIQualifier{\sun}{\ensuremath{\odot}}
\DeclareSIUnit\angstrom{\AA}
\DeclareSIUnit{\au}{au}
\DeclareSIUnit{\density}{\ensuremath{\mathnormal{\rho}}}
\DeclareSIUnit{\erg}{erg}
\DeclareSIUnit{\luminosity}{\ensuremath{\mathrm{L}}}
\DeclareSIUnit{\magnitude}{mag}
\DeclareSIUnit{\mass}{\ensuremath{\mathrm{M}}}
\DeclareSIUnit{\mas}{\milliarcsecond}
\DeclareSIUnit{\milliarcsecond}{mas}
\DeclareSIUnit{\parsec}{pc}
\DeclareSIUnit{\radius}{\ensuremath{\mathrm{R}}}
\DeclareSIUnit{\year}{yr}

\newcommand{\tablenotemark}[1]{\textsuperscript{\textit{#1}}}

\usepackage{xcolor}

\newcommand{\target}{TOI-5126}
\newcommand{\bPlanet}{TOI-5126 b}
\newcommand{\cPlanet}{TOI-5126 c}

\newcommand{\rsun}{\ensuremath{R_{\sun}}}
\newcommand{\msun}{\ensuremath{M_{\sun}}}
\newcommand{\lsun}{\ensuremath{L_{\sun}}}

\newcommand{\rearth}{\ensuremath{R_{\earth}}}
\newcommand{\mearth}{\ensuremath{M_{\earth}}}

\newcommand{\Searth}{\ensuremath{S_{\earth}}}

\newcommand{\mjup}{\ensuremath{M_{\rm Jup}}}

\newcommand{\teff}{\ensuremath{T_{\rm eff}}}
\newcommand{\teq}{\ensuremath{T_{\rm eq}}}
\newcommand{\logg}{\ensuremath{\log g}}
\newcommand{\feh}{[Fe/H]}
\newcommand{\vsini}{\ensuremath{v\sin I_\star}}

\newcommand{\rpl}{\ensuremath{R_{p}}}
\newcommand{\mpl}{\ensuremath{M_{p}}}

\newcommand{\kms}{\ensuremath{\rm km\,s^{-1}}}
\newcommand{\ms}{\ensuremath{\rm m\,s^{-1}}}

\newcommand{\rstar}{\ensuremath{R_{\star}}}
\newcommand{\mstar}{\ensuremath{M_{\star}}}
\newcommand{\lstar}{\ensuremath{L_{\star}}}

\newcommand{\teffstar}{\ensuremath{T_{\rm eff\star}}}
\newcommand{\rhostar}{\ensuremath{\rho_{\star}}}
\newcommand{\loggstar}{\ensuremath{\log{g_{\star}}}}
\newcommand{\arstar}{\ensuremath{a/\rstar}}

\newcommand{\ecosw}{\ensuremath{\sqrt{e}\cos\omega}}
\newcommand{\esinw}{\ensuremath{\sqrt{e}\sin\omega}}

\newcommand{\kepler}{{\it Kepler}}

\newcommand{\kt}{{\it K2}}
\newcommand{\tess}{{\it TESS}}
\newcommand{\cheops}{{\it CHEOPS}}

\newcommand{\lc}{light curve\ }
\newcommand{\lcs}{light curves\ }

\newcommand{\gcmc}{\ensuremath{\rm g\,cm^{-3}}}

\newcommand{\angstrom}{\textup{\AA}}

\newcommand{\Kb}{\ensuremath{K_b}}
\newcommand{\Kc}{\ensuremath{K_c}}
\newcommand{\Av}{\ensuremath{A_v}}

\newcommand{\ingress}{\ensuremath{t_{12}}}
\newcommand{\tduronethree}{\ensuremath{t_{13}}}

\newcommand{\Ic}{\ensuremath{I_{c}}}

\newcommand{\Insolc}{\ensuremath{S_c}}
\newcommand{\Teq}{\ensuremath{T_{\rm eq}}}
\newcommand{\rpb}{\ensuremath{R_{b}}}

\newcommand{\starTOIID}{\ensuremath{5126}}
\newcommand{\starTICID}{\ensuremath{27064468}}
\newcommand{\starGaiaID}{\ensuremath{623633615066248576}}
\newcommand{\starRA}{10:06:38.34}
\newcommand{\starDec}{+18:38:02.62}
\newcommand{\starRefEpoch}{\ensuremath{2016.0}} 
\newcommand{\starPMRA}{\ensuremath{-54.700\pm0.022}}
\newcommand{\starPMDec}{\ensuremath{-10.4043\pm0.016}}

\newcommand{\starParallax}{\ensuremath{6.224\pm0.022}}
\newcommand{\starTMag}{\ensuremath{9.5622\pm0.0064}}
\newcommand{\starGaiaMag}{\ensuremath{9.946\pm0.006}}
\newcommand{\starGaiaBPMag}{\ensuremath{10.222\pm0.013}}
\newcommand{\starGaiaRPMag}{\ensuremath{9.519\pm0.012}}
\newcommand{\starJMag}{\ensuremath{9.043\pm0.018}}
\newcommand{\starHMag}{\ensuremath{8.845\pm0.015}}
\newcommand{\starKMag}{\ensuremath{8.777\pm0.018}}
\newcommand{\starVMag}{\ensuremath{10.141\pm0.006}}
\newcommand{\starBMag}{\ensuremath{10.573\pm0.108}}
\newcommand{\starwoneMag}{\ensuremath{8.712\pm0.03}}
\newcommand{\starwtwoMag}{\ensuremath{8.743\pm0.03}}

\newcommand{\starDistance}{\ensuremath{161.1\pm1.2}}

\newcommand{\Kbbrown}{\ensuremath{1326}}
\newcommand{\Kcbrown}{\ensuremath{892}}

\newcommand{\tresteff}{\ensuremath{6217\pm50}}
\newcommand{\treslogg}{\ensuremath{4.3\pm0.1}}
\newcommand{\tresfeh}{\ensuremath{0.17\pm0.08}}
\newcommand{\tresvsini}{\ensuremath{14.8\pm0.5}}

\newcommand{\SEDteff}{\ensuremath{6150_{-100}^{+130}}}

\newcommand{\SEDlogg}{\ensuremath{4.31_{-0.23}^{+0.24}}}
\newcommand{\SEDfeh}{\ensuremath{0.18_{-0.13}^{+0.13}}}
\newcommand{\SEDrad}{\ensuremath{1.249_{-0.038}^{+0.040}}}

\newcommand{\Prot}{\ensuremath{4.602_{-0.067}^{+0.071}}}
\newcommand{\cTess}{\ensuremath{1.056_{-0.045}^{+0.023}}}
\newcommand{\cMcd}{\ensuremath{1.0001_{-0.0063}^{+0.0070}}}
\newcommand{\cTeid}{\ensuremath{0.9986_{-0.0060}^{+0.0075}}}
\newcommand{\cCtioOne}{\ensuremath{0.998_{-0.005}^{+0.010}}}
\newcommand{\cCtioTwo}{\ensuremath{0.9963_{-0.0044}^{+0.0084}}}
\newcommand{\cCheopsOne}{\ensuremath{1.075_{-0.040}^{+0.017}}}
\newcommand{\cCheopsTwo}{\ensuremath{1.011_{-0.035}^{+0.052}}}
\newcommand{\cCheopsThree}{\ensuremath{1.064_{-0.036}^{+0.025}}}

\newcommand{\starRho}{\ensuremath{0.922_{-0.079}^{+0.065}}} 
\newcommand{\starLogg}{\ensuremath{4.224_{-0.062}^{+0.051}}} 
\newcommand{\starMass}{\ensuremath{1.240_{-0.046}^{+0.050}}}
\newcommand{\starRadius}{\ensuremath{1.241_{-0.027}^{+0.032}}} 
\newcommand{\starLuminosity}{\ensuremath{1.99_{-0.17}^{+0.18}}} 
\newcommand{\starTeff}{\ensuremath{6150_{-130}^{+110}}} 
\newcommand{\starfeh}{\ensuremath{0.18_{-0.13}^{+0.13}}} 
\newcommand{\starAge}{\ensuremath{0.8_{-0.7}^{+2.8}}} 

\newcommand{\starTESSuOne}{\ensuremath{0.25_{-0.17}^{+0.17}}}
\newcommand{\starTESSuTwo}{\ensuremath{0.34_{-0.19}^{+0.25}}}
\newcommand{\starzsuOne}{\ensuremath{0.56_{-0.39}^{+0.33}}}
\newcommand{\starzsuTwo}{\ensuremath{0.26_{-0.39}^{+0.33}}}

\newcommand{\starcheopsuTwo}{\ensuremath{0.09_{-0.21}^{+0.29}}}

\newcommand{\bEpoch}{\ensuremath{2627.03862_{-0.00039}^{+0.00043}}}
\newcommand{\bPeriodapprox}{\ensuremath{5.46}}
\newcommand{\bPeriod}{\ensuremath{5.4588385_{-0.0000072}^{+0.0000070}}}
\newcommand{\bROR}{\ensuremath{0.03498_{-0.00060}^{+0.00065}}}

\newcommand{\bMassOtegi}{\ensuremath{21_{-7}^{+9}}}
\newcommand{\bImpactParameter}{\ensuremath{0.24_{-0.14}^{+0.10}}}
\newcommand{\bSemimajorAxis}{\ensuremath{0.06519_{-0.00082}^{+0.00087}}} 
\newcommand{\bAOR}{\ensuremath{11.32_{-0.33}^{+0.26}}} 
\newcommand{\bInclination}{\ensuremath{88.79_{-0.53}^{+0.72}}} 
\newcommand{\bRadius}{\ensuremath{4.74_{-0.14}^{+0.16}}} 
\newcommand{\bRadiusapprox}{\ensuremath{4.7}} 
\newcommand{\bPredictedK}{\ensuremath{6.6_{-2.5}^{+2.5}}} 
\newcommand{\bTeq}{{\ensuremath{{1442}_{-40}^{+46}}}} 
\newcommand{\bTSM}{\ensuremath{84_{-25}^{+53}}} 
\newcommand{\bRM}{\ensuremath{{11.70}_{-0.45}^{+0.41}}} 
\newcommand{\bIrradiation}{\ensuremath{467_{-39}^{+44}}} 
\newcommand{\bDuration}{\ensuremath{3.726_{-0.016}^{+0.017}}} 
\newcommand{\bIngressDuration}{\ensuremath{8.00_{-0.44}^{+0.54}}} 

\newcommand{\cEpoch}{\ensuremath{2613.7443_{-0.0029}^{+0.0049}}}

\newcommand{\cPeriod}{\ensuremath{17.8999_{-0.0013}^{+0.0018}}}
\newcommand{\cROR}{\ensuremath{0.02854_{-0.00090}^{+0.00080}}}

\newcommand{\cMassOtegi}{\ensuremath{18}_{-6}^{+8}}
\newcommand{\cImpactParameter}{\ensuremath{0.34_{-0.14}^{+0.10}}}
\newcommand{\cSemimajorAxis}{\ensuremath{0.1439_{-0.0018}^{+0.0019}}}
\newcommand{\cAOR}{\ensuremath{24.99_{-0.74}^{+0.57}}}
\newcommand{\cInclination}{\ensuremath{89.23_{-0.25}^{+0.32}}} 
\newcommand{\cRadius}{\ensuremath{3.86_{-0.16}^{+0.17}}}
\newcommand{\cPredictedK}{\ensuremath{3.8_{-1.5}^{+1.5}}} 
\newcommand{\cTeq}{\ensuremath{971_{-27}^{+31}}}
\newcommand{\cTSM}{\ensuremath{40_{-12}^{+27}}} 
\newcommand{\cRM}{\ensuremath{{7.52}_{-0.41}^{+0.43}}} 
\newcommand{\cIrradiation}{\ensuremath{96_{-8}^{+9}}} 
\newcommand{\cDuration}{\ensuremath{5.34_{-0.17}^{+0.14}}} 
\newcommand{\cIngressDuration}{\ensuremath{9.97_{-0.71}^{+0.85}}}

\newcommand{\bPeriodecc}{\ensuremath{5.4588382_{-0.0000070}^{+0.0000070}}}
\newcommand{\bEpochecc}{\ensuremath{2627.03862_{-0.00045}^{+0.00049}}}
\newcommand{\bDurationecc}{\ensuremath{3.72_{-0.17}^{+0.14}}}
\newcommand{\bIngressDurationecc}{\ensuremath{8.08_{-0.40}^{+0.62}}}
\newcommand{\bAORecc}{\ensuremath{11.26_{-0.36}^{+0.33}}}
\newcommand{\bRORecc}{\ensuremath{0.03511_{-0.00058}^{+0.00063}}}
\newcommand{\bEccentricityecc}{\ensuremath{0.09_{-0.07}^{+0.21}}}
\newcommand{\bOmegaecc}{\ensuremath{-10_{-130}^{+110}}}
\newcommand{\bImpactParameterecc}{\ensuremath{0.24_{-0.14}^{+0.15}}}
\newcommand{\bInclinationecc}{\ensuremath{88.82_{-0.82}^{+0.69}}}
\newcommand{\bRadiusecc}{\ensuremath{4.76_{-0.15}^{+0.17}}}
\newcommand{\bSemimajorAxisecc}{\ensuremath{0.0650_{-0.0009}^{+0.0011}}}
\newcommand{\bTeqecc}{\ensuremath{1445_{-46}^{+52}}}
\newcommand{\bIrradiationecc}{\ensuremath{471_{-44}^{+48}}}

\newcommand{\cPeriodecc}{\ensuremath{17.8999_{-0.0012}^{+0.0020}}}
\newcommand{\cEpochecc}{\ensuremath{2613.7440_{-0.0023}^{+0.0051}}}
\newcommand{\cDurationecc}{\ensuremath{5.25_{-0.57}^{+0.34}}}
\newcommand{\cIngressDurationecc}{\ensuremath{10.3_{-1.0}^{+2.5}}}
\newcommand{\cAORecc}{\ensuremath{24.85_{-0.79}^{+0.73}}}
\newcommand{\cRORecc}{\ensuremath{0.0287_{-0.0010}^{+0.0016}}}
\newcommand{\cEccentricityecc}{\ensuremath{0.21_{-0.16}^{+0.40}}}
\newcommand{\cOmegaecc}{\ensuremath{-20_{-120}^{+130}}}
\newcommand{\cImpactParameterecc}{\ensuremath{0.38_{-0.23}^{+0.20}}}
\newcommand{\cInclinationecc}{\ensuremath{89.13_{-0.49}^{+0.53}}}
\newcommand{\cRadiusecc}{\ensuremath{3.90_{-0.18}^{+0.26}}}
\newcommand{\cSemimajorAxisecc}{\ensuremath{0.1435_{-0.0020}^{+0.0025}}}
\newcommand{\cTeqecc}{\ensuremath{972_{-31}^{+35}}}
\newcommand{\cIrradiationecc}{\ensuremath{97_{-9}^{+10}}}

\newcommand{\bRadiusttv}{\ensuremath{4.67_{-0.13}^{+0.14}}}

\newcommand{\bDurationttv}{\ensuremath{3.686_{-0.026}^{+0.032}}}

\newcommand{\cRadiusttv}{\ensuremath{3.76_{-0.13}^{+0.14}}}

\newcommand{\cDurationttv}{\ensuremath{5.46_{-0.16}^{+0.10}}}

\newcommand{\bttvtest}{\ensuremath{0.27}}
\newcommand{\cttvtest}{\ensuremath{0.19}}

\title[\tess\ Discovery of Multi-Planets TOI-5126]{TOI-5126: A hot super-Neptune and warm Neptune pair discovered by \tess{} and \cheops}

\author[T.R. Fairnington et al.]{%
Tyler~R.~Fairnington%
\textsuperscript{\href{https://orcid.org/0000-0002-0692-7822}{\includegraphics[width=2.5mm]{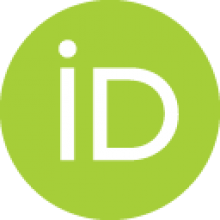}}}%
,\textsuperscript{1}%
\thanks{Email: tyler.fairnington@usq.edu.au}
Emma Nabbie%
\textsuperscript{\href{https://orcid.org/0000-0003-0571-2245}{\includegraphics[width=2.5mm]{orcid-ID.png}}}%
,\textsuperscript{1}
Chelsea~X.~Huang%
\textsuperscript{\href{https://orcid.org/0000-0003-0918-7484}{\includegraphics[width=2.5mm]{orcid-ID.png}}}%
,\textsuperscript{1}
\thanks{ARC DECRA Fellow}
George Zhou%
,\textsuperscript{1}
\thanks{ARC DECRA Fellow}
Orion Foo%
,\textsuperscript{2}
\newauthor
Sarah Millholland%
\textsuperscript{\href{https://orcid.org/0000-0003-3130-2282}{\includegraphics[width=2.5mm]{orcid-ID.png}}}%
,\textsuperscript{2}
Duncan Wright%
,\textsuperscript{1}
Alexandre~A.~Belinski%
\textsuperscript{\href{https://orcid.org/0000-0003-3469-0989}{\includegraphics[width=2.5mm]{orcid-ID.png}}}%
,\textsuperscript{3}
Allyson Bieryla%
,\textsuperscript{4}
David R. Ciardi %
\textsuperscript{\href{https://orcid.org/0000-0002-5741-3047}{\includegraphics[width=2.5mm]{orcid-ID.png}}}
,\textsuperscript{5}
\newauthor
Karen A.\ Collins%
\textsuperscript{\href{https://orcid.org/0000-0001-6588-9574}{\includegraphics[width=2.5mm]{orcid-ID.png}}}%
,\textsuperscript{4}
Kevin I.\ Collins%
\textsuperscript{\href{https://orcid.org/0000-0003-2781-3207}{\includegraphics[width=2.5mm]{orcid-ID.png}}}%
,\textsuperscript{6}
Mark Everett%
,\textsuperscript{7}
Steve~B.~Howell%
\textsuperscript{\href{https://orcid.org/0000-0002-2532-2853}{\includegraphics[width=2.5mm]{orcid-ID.png}}}%
,\textsuperscript{8}
Jack J. Lissauer%
\textsuperscript{\href{https://orcid.org/0000-0001-6513-1659}{\includegraphics[width=2.5mm]{orcid-ID.png}}}
,\textsuperscript{8}
\newauthor
Michael B. Lund%
\textsuperscript{\href{https://orcid.org/0000-0003-2527-1598}{\includegraphics[width=2.5mm]{orcid-ID.png}}}
,\textsuperscript{5}
Felipe Murgas%
\textsuperscript{\href{https://orcid.org/0000-0001-9087-1245}{\includegraphics[width=2.5mm]{orcid-ID.png}}}%
,\textsuperscript{9,10}
Enric Palle%
,\textsuperscript{9,10}
Samuel N. Quinn%
\textsuperscript{\href{https://orcid.org/0000-0002-8964-8377}{\includegraphics[width=2.5mm]{orcid-ID.png}}}%
,\textsuperscript{4}
Howard M. Relles%
,\textsuperscript{4}
\newauthor
Boris~S.~Safonov%
,\textsuperscript{3}
Richard P. Schwarz%
\textsuperscript{\href{https://orcid.org/0000-0001-8227-1020}{\includegraphics[width=2.5mm]{orcid-ID.png}}}%
,\textsuperscript{4}
Nicholas~J.~Scott%
\textsuperscript{\href{https://orcid.org/0000-0003-1038-9702}{\includegraphics[width=2.5mm]{orcid-ID.png}}}%
,\textsuperscript{8}
Gregor Srdoc%
,\textsuperscript{12}
George Ricker
\textsuperscript{\href{https://orcid.org/0000-0003-2058-6662}{\includegraphics[width=2.5mm]{orcid-ID.png}}}%
,\textsuperscript{2}
\newauthor
Roland Vanderspek
\textsuperscript{\href{https://orcid.org/0000-0001-6763-6562}{\includegraphics[width=2.5mm]{orcid-ID.png}}}%
,\textsuperscript{2}
Sara Seager
\textsuperscript{\href{https://orcid.org/0000-0002-6892-6948}{\includegraphics[width=2.5mm]{orcid-ID.png}}}%
,\textsuperscript{13,2,14}
David W. Latham
\textsuperscript{\href{https://orcid.org/0000-0001-9911-7388}{\includegraphics[width=2.5mm]{orcid-ID.png}}}%
,\textsuperscript{4}
Joshua W. Winn
\textsuperscript{\href{https://orcid.org/0000-0002-4265-047X}{\includegraphics[width=2.5mm]{orcid-ID.png}}}%
,\textsuperscript{15}\newauthor
Jon M. Jenkins
\textsuperscript{\href{https://orcid.org/0000-0002-4715-9460}{\includegraphics[width=2.5mm]{orcid-ID.png}}}%
,\textsuperscript{8}
Luke~G.~Bouma
\textsuperscript{\href{https://orcid.org/0000-0002-0514-5538}{\includegraphics[width=2.5mm]{orcid-ID.png}}}%
,\textsuperscript{16}\thanks{51 Pegasi b Fellow}
Avi Shporer
,\textsuperscript{2}
Eric B. Ting 
\textsuperscript{\href{https://orcid.org/0000-0002-8219-9505}{\includegraphics[width=2.5mm]{orcid-ID.png}}}%
,\textsuperscript{8}
Diana Dragomir
\textsuperscript{\href{https://orcid.org/0000-0003-2313-467X}{\includegraphics[width=2.5mm]{orcid-ID.png}}}%
,\textsuperscript{17}
\newauthor
Michelle Kunimoto
,\textsuperscript{2}
Nora L. Eisner
\textsuperscript{\href{https://orcid.org/ 0000-0002-0786-7307}{\includegraphics[width=2.5mm]{orcid-ID.png}}}%
,\textsuperscript{18, 15}
\\
\\
Affiliations are listed at the end of the paper
}

\begin{document}
\label{firstpage}
\pagerange{\pageref{firstpage}--\pageref{lastpage}}

\maketitle

\begin{abstract}
We present the confirmation of a hot super-Neptune with an exterior Neptune companion orbiting a bright (V = 10.1 mag) F-dwarf identified by the Transiting Exoplanet Survey Satellite (\tess). The two planets, observed in sectors 45, 46 and 48 of the \tess{} extended mission, are \bRadius \rearth{} and \cRadius \rearth{} with \bPeriod{} d and \cPeriod{} d orbital periods, respectively. We also obtained precise space based photometric follow-up of the system with ESA's CHaracterising ExOplanets Satellite (\cheops) to constrain the radius and ephemeris of \bPlanet. \bPlanet{} is located in the `hot Neptune Desert' and is an ideal candidate for follow-up transmission spectroscopy due to its high predicted equilibrium temperature (\Teq = \bTeq{} K) implying a cloud-free atmosphere. \cPlanet{} is a warm Neptune (\Teq = \cTeq{} K) also suitable for follow-up. Tentative transit timing variations (TTVs) have also been identified in analysis, suggesting the presence of at least one additional planet, however this signal may be caused by spot-crossing events, necessitating further precise photometric follow-up to confirm these signals.
\end{abstract}


\section{Introduction}
\label{sec:intro}

Super-Neptunes are a class of planets potentially bridging the gap between Neptune and Jovian planets\footnote{in the literature, these planets are also often referred as Sub-Saturns} (4-8 \rearth). Despite being of similar sizes, they exhibit a significant diversity in mass, almost uniformly distributed between 6 and 135 \mearth{} \citep{NEA, Petigura:2017}. Given the absence of such planets in the Solar system, studying the population provides extra insights on the versatility of the planet formation process around other stars. 

An archetypal example of the super-Neptune class is GJ 436 b \citep{Gillon2007}. It has a Neptune-like radius coupled with a relatively high density. It also has an inferred rocky element core and substantial H/He envelope \citep{Fortney2007}. Its characteristics appear to deviate from the theorised $\approx$ 5 \mearth{} threshold needed for runaway accretion \citep{Lee2016}, highlighting our lack of understanding about the formation channels from which this class originates. \kepler{} and the subsequent \kt{} mission \citep{kepler, k2} enabled the first population studies of these planets. \citet{Petigura:2017} proposed that super-Neptunes form in gas-depleted disks \citep{Lee2014} or from planet-planet mergers \citep{Petigura:2017}. Both of these formation scenarios lead to heavy elements in the planets atmosphere, making these theories testable with bright systems hosting super-Neptunes. 

However, the majority of confirmed super-Neptunes discovered by the \kepler{} and \kt{} missions orbit around relatively faint stars, posing challenges for detailed atmosphere characterisation. The all-sky \textit{Transiting Exoplanet Survey Satellite} (\tess, \citealt{tess}) mission aims to survey the entire  sky for candidates around some of the brightest nearby stars, providing the best opportunity to further our knowledge of super-Neptunes. \tess{} has already identified several unique super-Neptunes, including LTT 9779 b \citep{LTT9779b} and TOI-849 b \citep{TOI849}. Both reside in the `hot Neptune desert' \citep{Mazeh:2016}, creating further challenges to their formation and evolution models. Planets in this close-in highly irradiated period space are usually thought to be most susceptable to photoevaporation \citep{owen2012, jin2014, lopez2017} and tidal stripping \citep{tidal_effects2016, OwenLai2018}. LTT 9779 b is an ideal target for follow-up atmospheric studies, with a Spitzer phase curve already suggesting a high-metallicity atmosphere \citep{Crossfield2020}. Efforts to understand this population can be most fruitful via multiplanet systems, where comparative planetology is possible. 

In this paper, we present the discovery of a multi-transiting system around the F-dwarf (\teff $=$ \SEDteff{} K) \target. The system, first discovered by \tess, hosts a hot super-Neptune located in the `hot Neptune Desert', and an exterior warm Neptune. The brightness of \target{} and the high predicted equilibrium temperatures of planet b (\bTeq{} K) and c (\cTeq{} K) make these planets ideal for future follow-up, including atmospheric comparative planetology studies. In section \S\ref{sec:data}, we present the photometric, imaging  and spectroscopic observations that led to the discovery and confirmation of the \target{} system. Section \S\ref{sec:analysis} discusses the derived stellar parameters of \target{}, as well as the global modeling and analysis of the system. The results of our analysis are presented in section \S\ref{sec:discussion}, including validation of \bPlanet{} and \cPlanet{}, transit timing variations (TTVs), and future follow-up opportunities.

\section{Observations and Data analysis}
\label{sec:data}
\label{section:data}
\subsection{Photometry}
\subsubsection{TESS}
\label{sssec:tess}
{\target} (TIC 27064468) was observed by TESS nearly continuously on both the 2 min cadence target pixel stamps and the 10 min FFIs in sectors 45 (Camera 3 CCD 4) and 46 (Camera 2 CCD 2), with an additional sector (sector 48, Camera 1 CCD 4) a month later. The observations were taken from UT 2021-Nov-06 to UT 2022-Feb-26, for a total of $\approx$76 days. 

In sector 46, systematic errors were observed in the first 13 hours of data due to a pointing adjustment \footnote{\url{https://archive.stsci.edu/missions/tess/doc/tess_drn/tess\_sector\_46\_drn66\_v02.pdf}}. To mitigate these errors, we utilised quality flags to mask out scattered light and other systematic effects in the \lcs{}. 

The \lcs{} of \target{} were processed by both the \tess{} Science Processing Operations Center (SPOC; \citealt{spoc}), located at NASA Ames Research Center. The SPOC conducted a transit search of Sector 45 2-min cadence data on UT 2021-Jan-09 with an adaptive, noise-compensating matched filter \citep{Jenkins:2002, Jenkins:2010, Jenkins:2020}, producing a Threshold Crossing Event (TCE) for which an initial limb-darkened transit model was fitted \citep{Li:2019} and a suite of diagnostic tests were conducted to help make or break the planetary nature of the signal \citep{Twicken:2018}. Transit signatures were also detected in a search of Full Frame Image (FFI) 10-min cadence data by the Quick Look Pipeline (QLP) at MIT \citep{qlp}. The TESS Science Office (TSO) reviewed the vetting information and issued an alert for \bPlanet{} on UT 2022-Jan-25 based and an alert for \cPlanet{} on UT 2022-Feb-08 \citep{toi} based on QLP data. The signals were repeatedly recovered as additional observations were made including in both Sectors 46 and 48, as well as SPOC multi-sector searches (42-46, 14-50). The transit signatures again passed all the diagnostic tests presented in the Data Validation reports. Specifically, difference images from the Sectors 42-46 data located the catalog location of \bPlanet{} and \cPlanet{} within \ensuremath{9.0\pm{5.0}} \arcsec{} and \ensuremath{5.2\pm{5.8}} \arcsec{} of the transit source. The latest SPOC results captured signals for planets b (Signal-to-Noise (SNR): 26.1; Multiple Event Statistic (MES): 19.9) and c (SNR: 10.4; MES: 9.9). The transit events of \bPlanet{} were also identified by the Planet Hunters \tess{} citizen science project \citep{Eisner:2020} and uploaded to exoFOP as communitiy objects of interest (CTOI) on UT 2022-Jan-12.

\begin{figure*}
\centering
    \includegraphics[width=15cm]{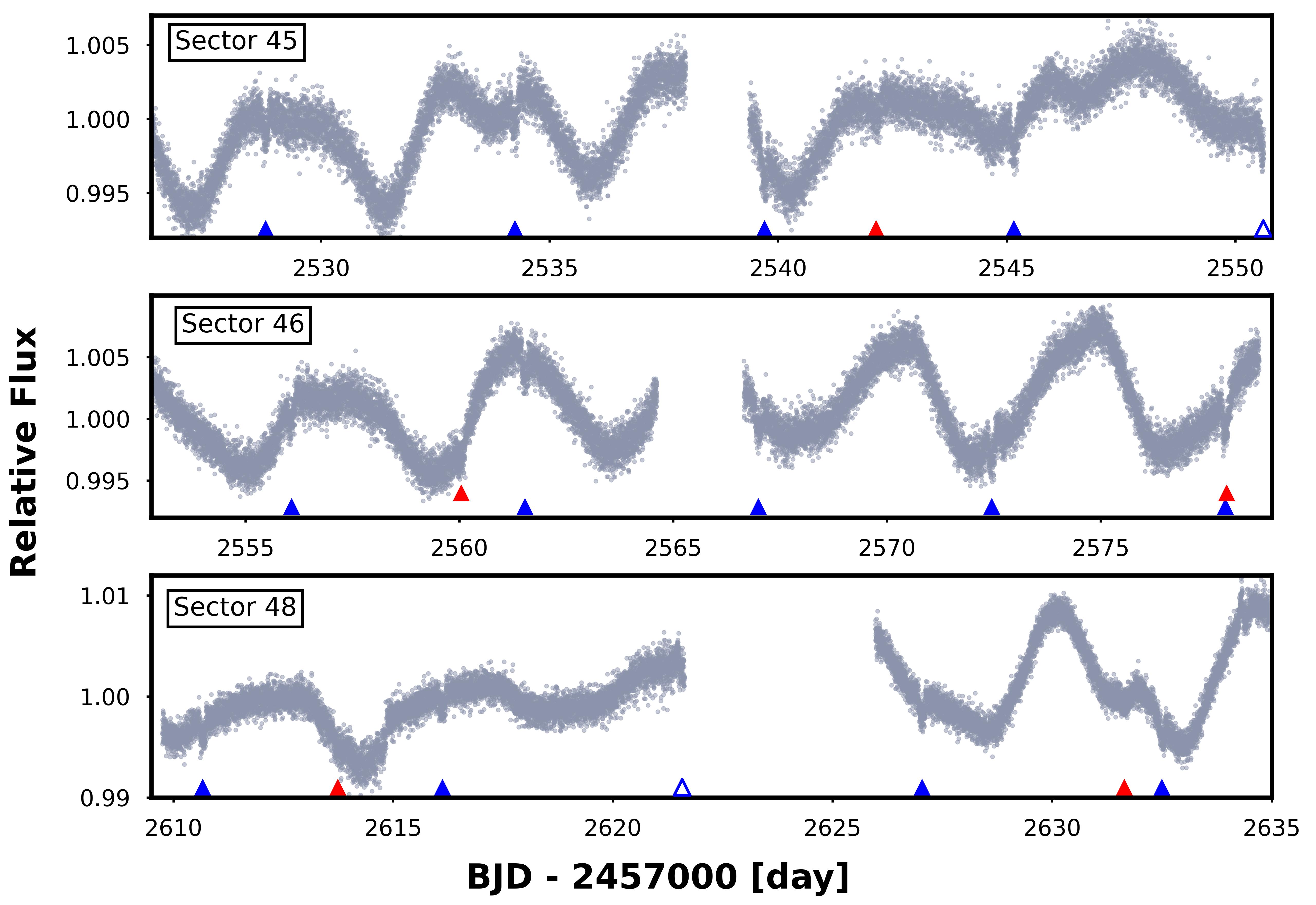}
\caption{\tess{} SPOC SAP \lc{} of \target{} from all observed sectors. Blue triangles mark each transit of \bPlanet{}, with red arrows for transits of \cPlanet{}. Empty triangles indicate the transits not used in global modeling. \label{fig:raw_lc}}
\end{figure*}

For our analysis, a modification of the deblended SPOC Simple Aperture Photometry (SAP) \lcs{} \citep{Twicken:2010, Morris:2020} were used in modeling (see Figure \ref{fig:raw_lc}). Stellar variability was accounted for by simultaneously fitting for a polynomial trend around each transit as part of our global analysis. Two transits were manually excluded due to them occurring near a data gap. Our ephemeris and radius precision did not noticeably suffer from these excluded transits. The details of the global fit will be presented in section \S\ref{ssec:globalfit}. 

\begin{figure*}
\centering
    \includegraphics[width=15cm]{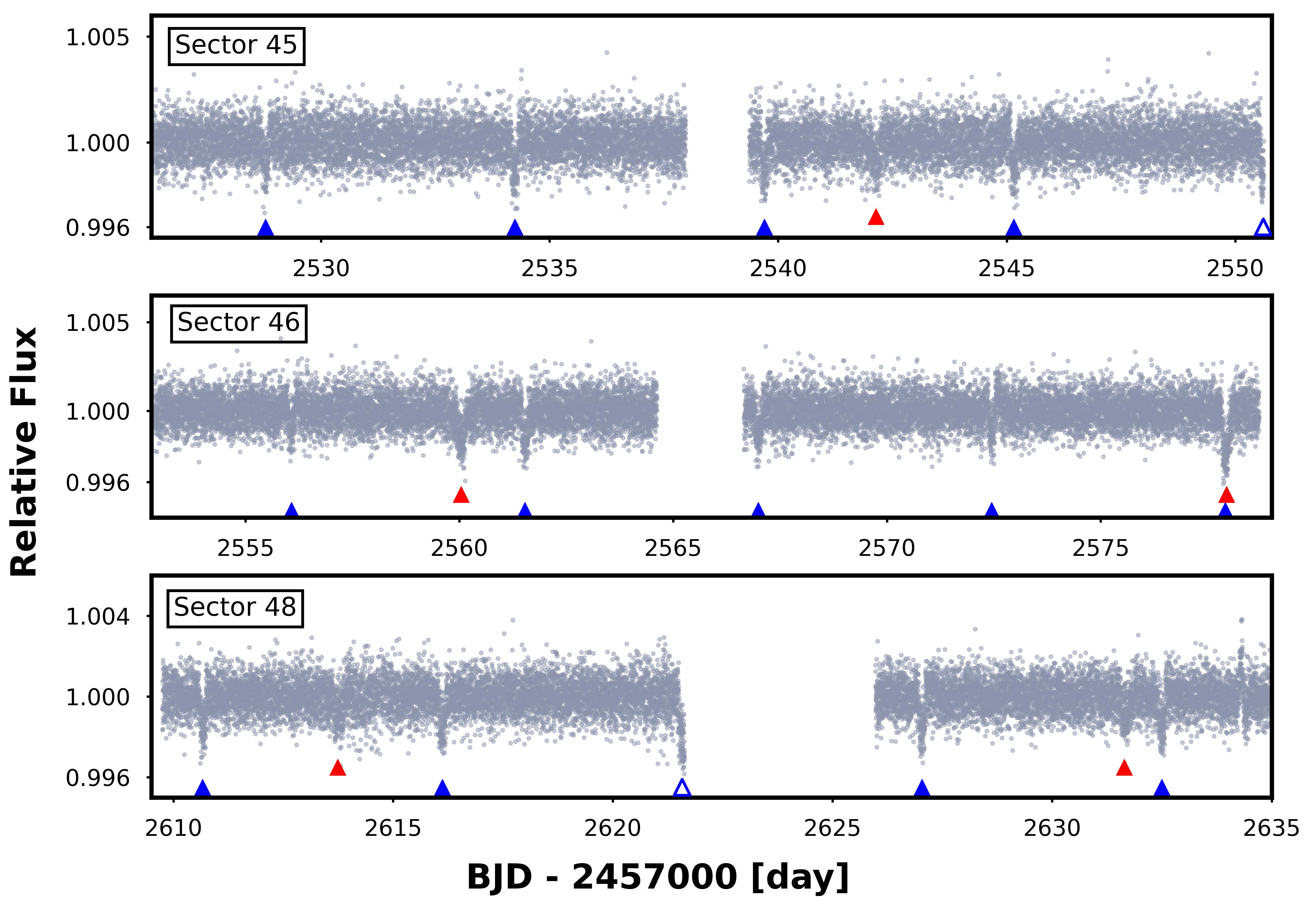}
\caption{Detrended \tess{} \lc{} of \target{} from all observed sectors. The light curves are detrended using an iterative B-spline method, and are not used in the global modeling that lead to the final constraint of the planet parameters. Blue triangles mark each transit of \bPlanet{}, with red arrows for transits of \cPlanet{}. Empty triangles indicate the transits not used in global modeling. \label{fig:detrended}}
\end{figure*}

\begin{figure*}
\includegraphics[width=\textwidth,height=\textheight,keepaspectratio]{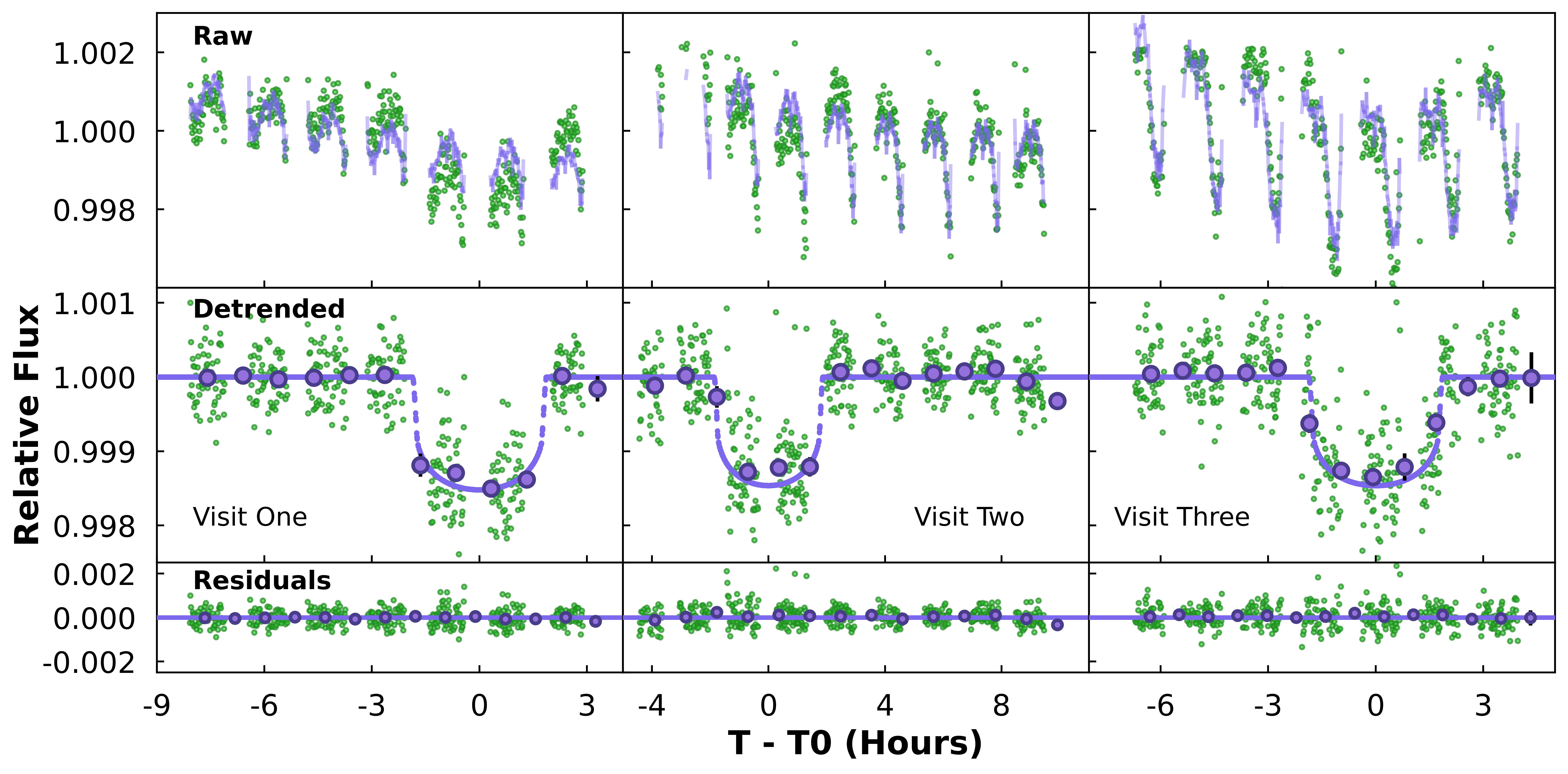}
\caption{\textbf{Upper}: Raw normalised \cheops{} \lcs{} indicated by green points with the best-fit trend model plotted in purple. \textbf{Middle}: Detrended \cheops{} \lcs{} shown in green. Binned points are denoted as purple markers and the best-fit transit model is overplotted as a purple line. \textbf{Lower}: Residuals of the detrended \lcs{} in green, with binned averaged points shown in purple.}
\label{fig:cheopsplots}
\end{figure*}

\begin{figure*}
\centering
\includegraphics[width=\textwidth,height=\textheight,keepaspectratio]{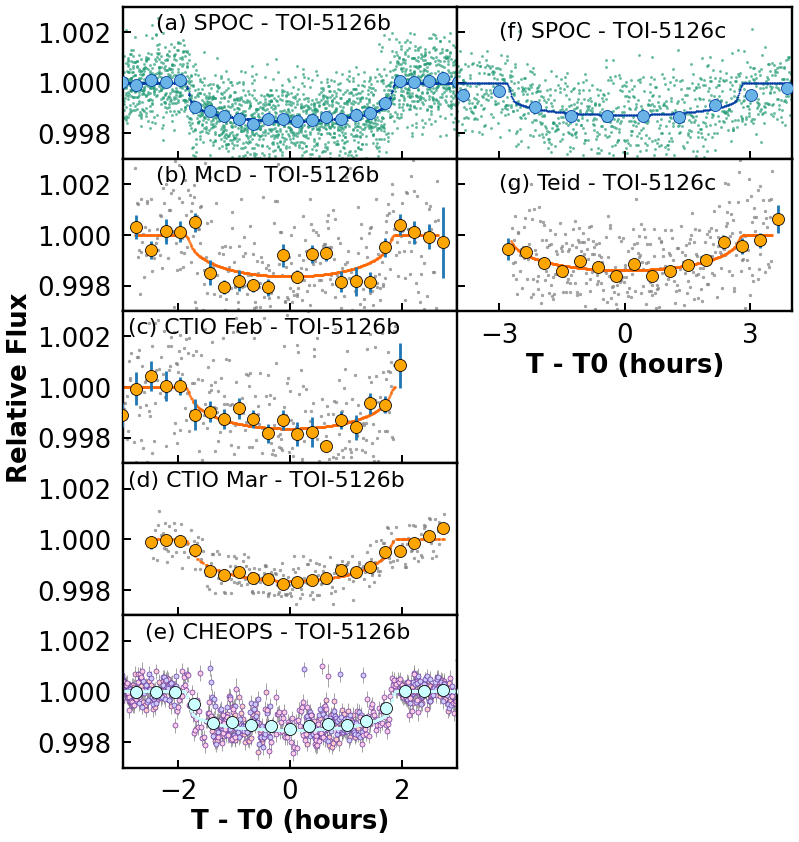}
\caption{Phase-folded \lcs{} of \bPlanet{} and \cPlanet. \tess{} SPOC \lcs{} are noted by green points, with binned data in blue and the best-fit transit model line in blue. LCOGT data is represented by gray points with orange binned points and errorsbars in blue. The transit model is also denoted as an orange line. The \cheops{} phase folded \lc{} is combined with all three visits \bPlanet, where the purple, pink and blue points represent each visit respectively. The binned data is in cyan, with the transit model also a cyan line. \textbf{(a)} \tess{} SPOC \bPlanet. \textbf{(b)} McD \bPlanet. \textbf{(c)} CTIO 2022 \bPlanet.  \textbf{(d)} CTIO 2023 \bPlanet.  \textbf{(e)} CHEOPS combined visits of \bPlanet. \textbf{(f)} \tess{} SPOC \cPlanet.  \textbf{(g)} Teid \cPlanet.}
\label{fig:phaseplot}
\end{figure*}

\subsubsection{CHEOPS}
\label{sssec:cheops}

To refine the radius and ephemeris of \bPlanet, we obtained space-based photometric observations with \cheops{} \citep{cheops}. Observations were attained through the \cheops{} Guest Observers Programme (AO3-023). A first observation of \bPlanet{} occurred on 2022-02-16 21:44 UT for a duration of 11.72 hours. The preliminary data was processed through the \cheops{} Data Reduction Pipeline (version 13.1; \citealt{cheopspipeline}) on 2023-Feb-17 18:46 UT. The visit consisted of seven \cheops{} orbits, with a four orbit pre-ingress and one orbit post-egress out-of-transit baseline. An additional visit of \bPlanet{} occurred on 2023-Feb-27 23:55 UT in an attempt to identify an ingress/egress with high efficiency, resulting in an longer 14.14 hour observation, with two orbits pre-ingress and five orbits post-egress. We obtained a third visit of \bPlanet{} on 2023-Mar-27 04:44 where we successfully caught both the ingress and egress over an observation duration of 10.96 hours. The full details of observations are listed in Table \ref{tab:cheops}. 

The raw \cheops{} \lcs{} were detrended simultaneously with the transit modeling to best incorporate and propagate the uncertainties associated with the instrument into our final system parameters. Observations with poor data quality flags, and those that fail an iterative 5-$\sigma$ outlier rejection, were rejected from subsequent analyses. To correct for the \cheops{} systematics, we followed the trend model in the \textsc{Pycheops} package \citep{pycheops}. This model fits for spacecraft and environmental variations, including the roll angle, $\phi$, the background flux, \textsc{bg}, the contamination in the aperture, \textsc{contam}, smearing correction, \textsc{smear} and the x and y point spread function (PSF) centroid coordinates, \textsc{dx $\&$ dy}. The raw \cheops{} \lcs{} were detrended by fitting a least-squares model to the light curve residuals, which provided us with the best parameters for the trend model. These parameters were then applied to the trend function and subtracted from the raw flux to provide us with the detrended \cheops{} \lcs, seen in Figure \ref{fig:cheopsplots}. 

\begin{table*}
\caption{
List of CHEOPS Observations
}
\label{tab:cheops}
\renewcommand{\arraystretch}{1.25}
\begin{tabular}{llllllll}
\toprule
Planet & Start Date [UT] & Duration [hours] & Valid Points [\#] & File Key & Efficiency [\%] & Exp. Time [s] \\

\midrule
b & 2023-02-16 21:44 & 11.72 & 422 & CH\_PR230023\_TG000301\_V0200 & 59.20 & 60.0 \\
b & 2023-02-27 23:55 & 14.14 & 549 &  CH\_PR230023\_TG000302\_V0200 & 64.60 & 60.0 \\
b & 2023-03-27 04:44 & 10.96 & 458 &  CH\_PR230023\_TG000501\_V0200 & 65.30 & 60.0 \\ 

\bottomrule
\addlinespace[4pt]
\end{tabular}

\raggedright

\end{table*}

\subsubsection{Ground based photometric follow up}
\label{sssec:groundbasedphotometry}
Due to the coarse angular resolution of both \tess{} and \cheops, eclipsing binary stars within their apertures may mimic a transit event on the target star. To confirm the on-target detection of \target, we obtained ground-based photometric observations of both planets. We used the {\tt TESS Transit Finder}, which is a customized version of the {\tt Tapir} software package \citep{Jensen:2013}, to schedule our transit observations. \bPlanet{} was observed in the $z_s$ filter by two 1.0 m Las Cumbres Observatory Global Network (LCOGT; \citealt{brown2013}) telescopes simultaneously at the McDonald Observatory (McD) in Fort Davis, Texas (2022-Mar-10 UT). The transit event was detected with a depth of $\sim$1.8 ppt. The transit occured $\sim$12 minutes late (possibly caused by TTVs, see Section \S \ref{sssec:ttv_lc}) and the field was cleared of Nearby Eclipsing Binaries in a 5.8\arcsec{} aperture. An ingress was also observed by the 1.0 m telescope at the Cerro Tololo Inter-American Observatory (CTIO) in Cerro Tololo, Chile (2022-Apr-01 UT), however, this was not used in subsequent analysis and modeling of the system. \cPlanet{} was also observed with two 1.0 m LCOGT telescopes simultaneously at the Teide Observatory (Teid) in Tenerife, Spain on 2022-Apr-01 UT. A transit event was detected ($\sim$0.6 ppt) $\sim$57 minutes late after double-detrending with a variable aperture of 4.7\arcsec{} and 3.1\arcsec. Two additional full transits of \bPlanet{} were obtained by the 1.0 m telescope at CTIO (2023-Feb-17 UT \& 2023-Mar-15 UT). Both full coverages resulted in $\sim$1.2 ppt detections in an uncontaminated 4.7\arcsec{} aperture. Table \ref{tab:sg1} provides the details of observation.

The images were calibrated by the standard LCOGT {\tt BANZAI} pipeline \citep{McCully:2018} and differential photometric data were extracted using {\tt AstroImageJ} \citep{Collins:2017}. We opted to perform our own detrending of all LCOGT \lcs{} simultaneous to the global fit. The trend function for the \lcs{} were represented by a second-order polynomial, with an additional term counting for the full width half maximum of the PSF. As with the CHEOPS trend model, the coefficients of the detrending vectors were derived from a least squares fit that was performed at each iteration of the global modeling process.

\begin{table*}
\caption{
 SG1 Follow-Up Observations 
}
 \label{tab:sg1}
\renewcommand{\arraystretch}{1.25}
\begin{tabular}{lllllll}
\toprule
Target & Instrument & Date (UT) & Filter & Aperture & Observing Notes\\

\midrule

\cPlanet & LCO-Teid 1.0m & 2022-03-10 & $z_s$ & 5.8\arcsec & Full coverage, simultaneous observation \tablenotemark{a} \\
\cPlanet & LCO-Teid 1.0m & 2022-03-10 & $z_s$& 4.7\arcsec & Full coverage, simultaneous observation \tablenotemark{a} \\
\bPlanet & LCO-McD 1.0m & 2022-04-01 & $z_s$ & 5.8\arcsec &  Full coverage, simultaneous observation \tablenotemark{b}\\
\bPlanet & LCO-McD 1.0m  & 2022-04-01 & $z_s$ & 4.7\arcsec & Full coverage, simultaneous observation \tablenotemark{b} \\
\bPlanet & LCO-CTIO 1.0m  & 2022-04-01 & $z_s$ & 3.9\arcsec & Ingress only, not included in global analysis \\
\bPlanet & LCO-CTIO 1.0m  & 2023-02-17 & $z_s$ & 4.7\arcsec & Full coverage \\
\bPlanet & LCO-CTIO 1.0m & 2023-03-15 & $z_s$ & 4.7\arcsec & Full coverage \\
\bottomrule
\addlinespace[4pt]
\end{tabular}
\flushleft
\tablenotemark{a}Transit observed simultaneously by two distinct telescopes at Teid. \\

\tablenotemark{b}Transit observed simultaneously by two distinct telescopes at McD.

\end{table*}

\subsection{Reconnaissance Spectroscopy}
\label{ssec:tres}

\begin{figure*}
\includegraphics[width=\textwidth,height=\textheight,keepaspectratio]{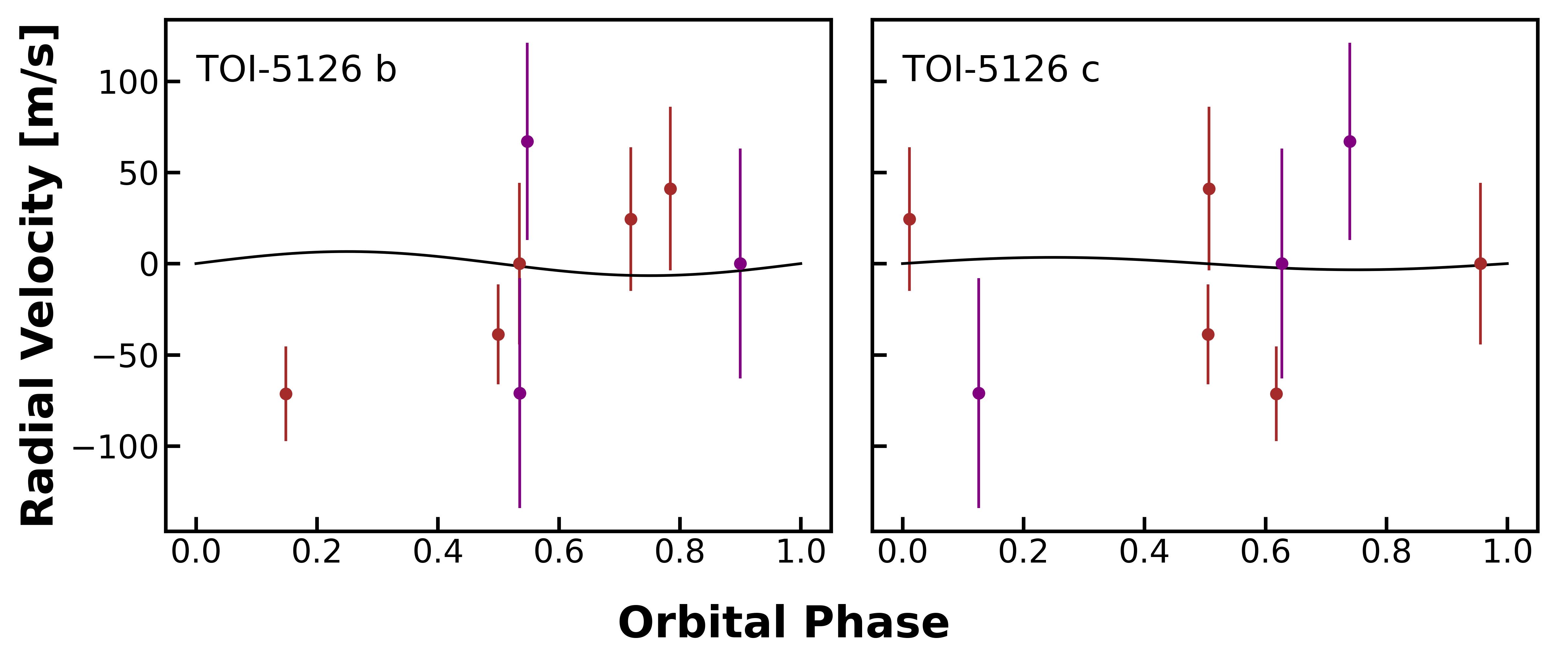}
\caption{\textbf{Left}: Radial velocities against the orbital phase of \bPlanet{} with the notional K semi-amplitude curve based off of predicted mass in Section \ref{ssec:followup}. TRES points are represented by purple points, while CHIRON data is in brown. \textbf{Right}: Radial velocities versus orbital phase for \cPlanet. with notional K semi-amplitude curve drawn from predicted mass.}
\label{fig:rvs}
\end{figure*}

We obtained a series of spectroscopic observations to check for contaminating spectroscopic blend scenarios, and to better constrain the spectroscopic atmospheric parameters of the target star. These reconnaissance spectroscopic observations are detailed in Table~\ref{tab:recon} and seen in Figure~\ref{fig:rvs}.  

We obtained three spectra of \target{} using the Tillinghast Reflector Echelle Spectrograph (TRES; \citealt{TRES}) on the 1.5 m telescope at the Fred Lawrence Whipple Observatory (Mt. Hopkins, Arizona). TRES has a resolving power of R $\simeq$ 44,000, with a wavelength coverage of 3850 to 9096 \angstrom. The observations were obtained with an average exposure SNR per resolution element of 38 at 5110 \angstrom. Radial velocities are derived from a multi-order cross correlation analysis as per \citet{2012ApJ...745...80Q}, and spectroscopic parameters were derived via the Spectroscopic Parameter Classification tool  \citep{Buchhave2010}. Observations yielded stellar parameters for effective temperature (\teff{} = \tresteff{} K), stellar surface gravity (\logg{} = \treslogg{} cgs),  metallicity (\feh{} = \tresfeh{} dex) and projected rotation velocity (\vsini= \tresvsini{} \kms). 

In addition, we also obtained five observations of \target{} with the CHIRON high resolution spectrograph, located on the 1.5\,m SMARTS telescope at Cerro Tololo Inter-American Observatory (CTIO), Chile \citep{2013PASP..125.1336T}. Observations were obtained using the fiber fed image slicer, yielding a resolution of $R\sim 80,000$. The spectra were extracted and reduced via the official CHIRON pipeline \citep{2021AJ....162..176P}. Radial velocities were determined from the modeling of their least-squares deconvolution profiles, generated against a non rotating synthetic template \citep{2021AJ....161....2Z}. In addition, we also model the rotational broadening of the target as a combination of the rotational, instrument, and macroturbulent broadening kernels. We derive broadening velocities of $v\sin I_\star = 13.2 \pm 0.5 \,\mathrm{km\,s}^{-1}$ and $v_\mathrm{macro} = 6.8\pm0.5\,\mathrm{km\,s}^{-1}$ for \target{}. 

We find no large radial velocity variations at the $>50\,\mathrm{m\,s}^{-1}$ level, or line profile deviations, as is expected for a system of small planets. The results indicate no obvious spectroscopic blending are present in the system. 

\begin{table*}
\caption{SG2 Reconnaissance Spectroscopy}
\label{tab:recon} 
\renewcommand{\arraystretch}{1.25}
\begin{tabular}{cccccccccc}
\toprule
Telescope & Instrument & Date (UT) & SNR & BJD & RV (kms$^{-1}$) & $\sigma_{RV}$ \\

\midrule

FLWO (1.5 m)  & TRES & 2022-03-04 & 40 & 2459642.86967 & 0.000\tablenotemark{a} & 0.063\\
FLWO (1.5 m)  & TRES & 2022-03-13 & 40 & 2459651.79751 & -0.071\tablenotemark{a} & 0.063\\
FLWO (1.5 m)  & TRES & 2022-03-24 & 40 & 2459662.78279 & 0.067\tablenotemark{a} & 0.054\\
CTIO (1.5 m) & CHIRON & 2022-03-02 & 200 & 2459640.68363 & 11.074 & 0.041\\
CTIO (1.5 m) & CHIRON & 2022-03-20 & 200 & 2459658.61297 & 11.154 & 0.050\\
CTIO (1.5 m) & CHIRON & 2022-03-22 & 210 & 2459660.60426 & 11.041 & 0.044\\
CTIO (1.5 m) & CHIRON & 2022-04-15 & 170 & 2459684.54751 & 11.113 & 0.056\\
CTIO (1.5 m) & CHIRON & 2022-04-16 & 180 & 2459685.55202 & 11.137 & 0.044\\

\bottomrule
\addlinespace[4pt]
\end{tabular}
\flushleft
\tablenotemark{a}Relative velocities from TRES self cross correlation analysis

\end{table*}

\subsection{High resolution imaging} 
\label{ssec:imaging}

As part of our standard process for validating transiting exoplanets to assess the the possible contamination of bound or unbound companions on the derived planetary radii \citep{ciardi2015}, we observed TOI-5126 with high-resolution near-infrared adaptive optics (AO) imaging at Palomar Observatory and in the optical with speckle imaging at WIYN and SAI.  The infrared observations provide the deepest sensitivities to faint companions while the optical speckle observations provide the highest resolution imaging making the two techniques complementary.  Combined with Gaia, the high resolution imaging observations find no evidence for additional stellar companions within the system.

\subsubsection{Palomar 5m/PHARO}
\label{sssec:palomar}

The Palomar Observatory observations were made with the PHARO instrument \citep{hayward2001} behind the natural guide star AO system P3K \citep{dekany2013} on 2022~Feb~13 UT in a standard 5-point quincunx dither pattern with steps of 5\arcsec\ in the narrow-band $Br-\gamma$ filter $(\lambda_o = 2.1686; \Delta\lambda = 0.0326~\mu$m).  Each dither position was observed three times, offset in position from each other by 0.5\arcsec\ for a total of 15 frames; with an integration time of 5.7 seconds per frame, the total on-source time was 85.5 seconds. PHARO has a pixel scale of $0.025\arcsec$ per pixel for a total field of view of $\sim25\arcsec$. The science frames were flat-fielded and sky-subtracted.  The reduced science frames were combined into a single combined image with a final resolution of 0.97\arcsec FWHM.

To within the limits of the AO observations, no stellar companions were detected. The sensitivities of the final combined AO image were determined by injecting simulated sources azimuthally around the primary target every $20^\circ $ at separations of integer multiples of the central source's FWHM \citep{furlan2017, lund2020}. The brightness of each injected source was scaled until standard aperture photometry detected it with $5-\sigma $ significance. The resulting brightness of the injected sources relative to TOI~5126 set the contrast limits at that injection location. The final $5-\sigma $ limit at each separation was determined from the average of all of the determined limits at that separation and the uncertainty on the limit was set by the rms dispersion of the azimuthal slices at a given radial distance. Figure \ref{fig:directimage} shows no stellar companion with a difference in magnitude ($\Delta$mag) of 5.4 at a separation $\sim$0.3\arcsec{}, and all companions ruled out in a contrast of $\Delta$mag of 9 from $\sim$2\arcsec{} onwards. No sources are present in the field-of-view with an upper limit on the angular separation of 4\arcsec.

\subsubsection{WIYN 3.5m/NESSI}
\label{sssec:wiyn}

\begin{figure*}
\centering
\includegraphics[width=.5\textwidth]{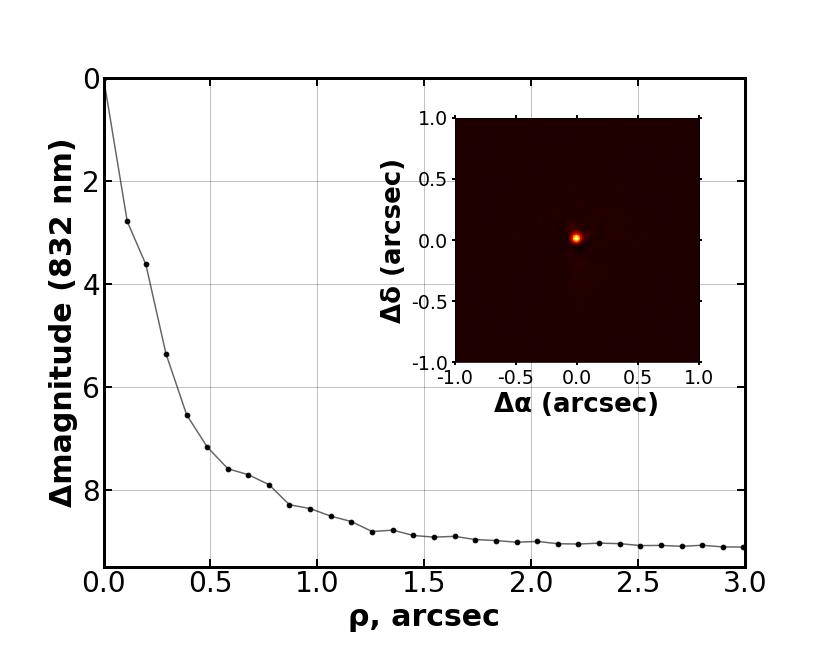}\hfill
\includegraphics[width=.5\textwidth]{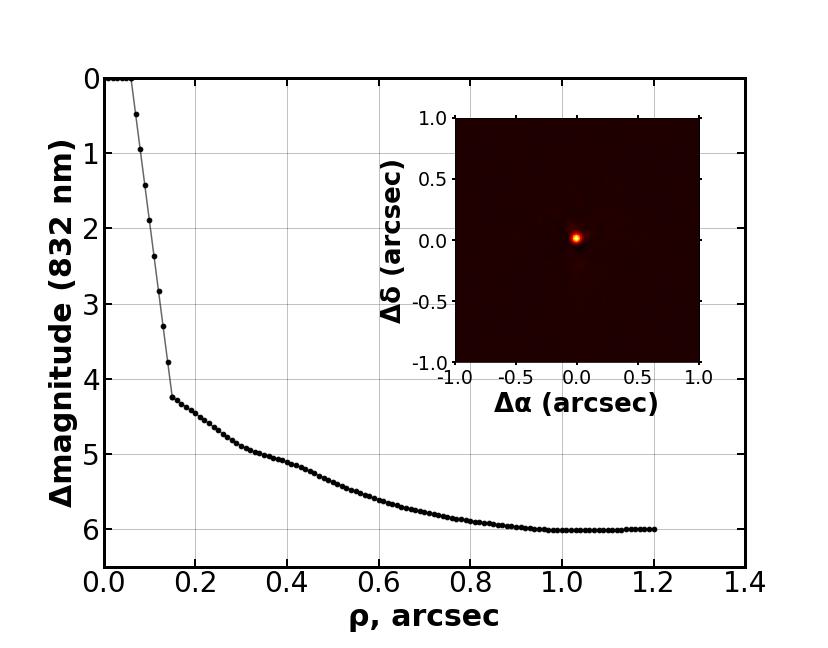}\hfill

\caption{{\textit{Left}: Palomar 5.0 m companion sensitivity curve. The black points represent the 5-$\sigma$ limits. The inset image is of the primary target showing no additional close-in companions. \textit{Right}: Contrast curve of \target{} with the NESSI instrument on the WIYN 3.5 m telescope, reaching a constrast of $\Delta$ 6 mag from 0.96\arcsec.}}

\label{fig:directimage}

\end{figure*}

Further imaging observations were made using the NN-Explore Exoplanet Stellar Speckle Imager (NESSI; \citealt{nessi}) on the WIYN 3.5 m telescope at the Kitt Peak National Observatory (KPNO) in Arizona. The observation and data reduction of NESSI can be found in \citet{nessireduction}. Images were taken at 832nm on UT 2022-Apr-18. The 5-$\sigma$ sensitivity curve, imaged in Figure \ref{fig:directimage}, outline a $\Delta$m of 6 from 0.96\arcsec{} onwards. The inset image (1.2\arcsec{} angular separation) also shows no stellar neighbours within the $\Delta$m of 6. 

\subsubsection{SAI 2.5m/Speckle Polarimeter}
\label{sssec:sai}

A third direct image of \target{} was taken using the Speckle Polarimeter on the 2.5 m Sternberg Astronomical Institute (SAI; \citealt{SAI2.5m}). Observations were made in the \Ic{} filter ($\approx 660nm - 825nm$). Observations were in agreement with the prior direct imaging observations, reaching a contrast of $\Delta$m of 7 from 1\arcsec. 

\subsubsection{Gaia Assessment}
In addition to the high resolution imaging, we have utlized Gaia to identify any wide stellar companions that may be bound members of the system.  Typically, these stars are already in the TESS Input Catalog and their flux dilution to the transit has already been accounted for in the transit fits and associated derived parameters.  Based upon similar parallaxes and proper motions \citep[e.g.,][]{mugrauer2020,mugrauer2021,mugrauer2022}, there are no additional widely separated companions identified by Gaia.

Additionally, the Gaia DR3 astrometry provides additional information on the possibility of inner companions that may have gone undetected by either Gaia or the high resolution imaging. The Gaia Renormalised Unit Weight Error (RUWE) is a metric, similar to a reduced $\chi^2$, where values that are $\lesssim 1.4$  indicate that the Gaia astrometric solution is consistent with the star being single whereas RUWE values $\gtrsim 1.4$ may indicate an astrometric excess noise, possibily caused the presence of an unseen companion \citep[e.g., ][]{ziegler2020}.  \target{} has a Gaia DR3 RUWE value of 1.16 indicating that the astrometric fits are consistent with the single star model. In addition the GAIA DR3 non-single-star flag is no present for the system, implying further that this is a single star system. 

\section{Analysis}
\label{sec:analysis}
\label{section:analysis}
\subsection{Stellar Parameters}
\label{ssec:stellarparameters}
Stellar parameters for \target{} were obtained by fitting the star's broadband spectral energy distribution (SED) to stellar atmosphere models using the Python package \textsc{astroARIADNE} \citep{ariadne}. We used TRES spectroscopic data to define priors on the stellar effective temperature (\teff) (with its uncertainty inflated to 150\,K) and metallicity (Fe/H), while distance constraints were derived from Gaia DR3 parallax \citep{gaiacollaboration2022}. We used the default prior for Extinction (\Av) which constrains an upper limit using the SFD galactic dust maps from \citet{Avdustmap}.

\textsc{astroARIADNE} uses a Bayesian Model Averaging (BMA) approach, fitting multiple models \footnote{The models used for this specific fit are: Phoenixv2 \citep{Husser:2013},  BT-Settl \citep{Allard:2012},
	BT-NextGen \citep{Allard:2012},
	BT-Cond \citep{Hauschildt:1999,Allard:2012},
	and Kurucz93 \citep{Kurucz:1993}} accounts for model-specific systematic biases. Each model is weighed individually for a combined weighted and averaged posterior probability. This is then fed into the isochrone fitting package. In SED fitting, we used Tycho-2 \citep{Tycho2} B and V, Gaia DR3 \citep{fabricius2021} G, Bp and Rp, Two-Micron All-Sky Survey (2MASS) \citep{2MASS} J, H, K, and Wide-field Infrared Survey Explorer (WISE) \citep{WISE} $W_1$ and $W_2$ bands, with an uncertainties floor applied to all bands based on \citet{Eastman:2019}. 
The weighted average results obtained are \teff{} $=$ \SEDteff{} K, \logg{} $=$ \SEDlogg{} cgs, \rstar{} $=$ \SEDrad{}\, 
\rsun{} 
and \feh{} $=$ \SEDfeh{} dex. 

To identify the rotation period of \target, we passed the raw deblended SPOC \lc{} through a Lomb-Scargle periodogram. The highest power of the entire \tess{} set of observations was constrained to be \Prot{} days. We used the FWHM of the highest power to derive our uncertainties. We also measured the Lomb-Scargle peaks of each sector separately, finding a scatter in the derived rotation period of $\sigma = 0.066$ days, consistent with the uncertainty estimated from the peak FWHM. However, the actual uncertainty of the rotation period may be much larger and is difficult to quantify with the short observational baseline available. The periodogram can be seen in Figure \ref{fig:periodogram}. Given the system has a measured \vsini{} of $\sim 14$ km s$^{-1}$, the inclination of the star is consistent with 90 degrees.

\begin{figure}
\includegraphics[width=.49\textwidth]{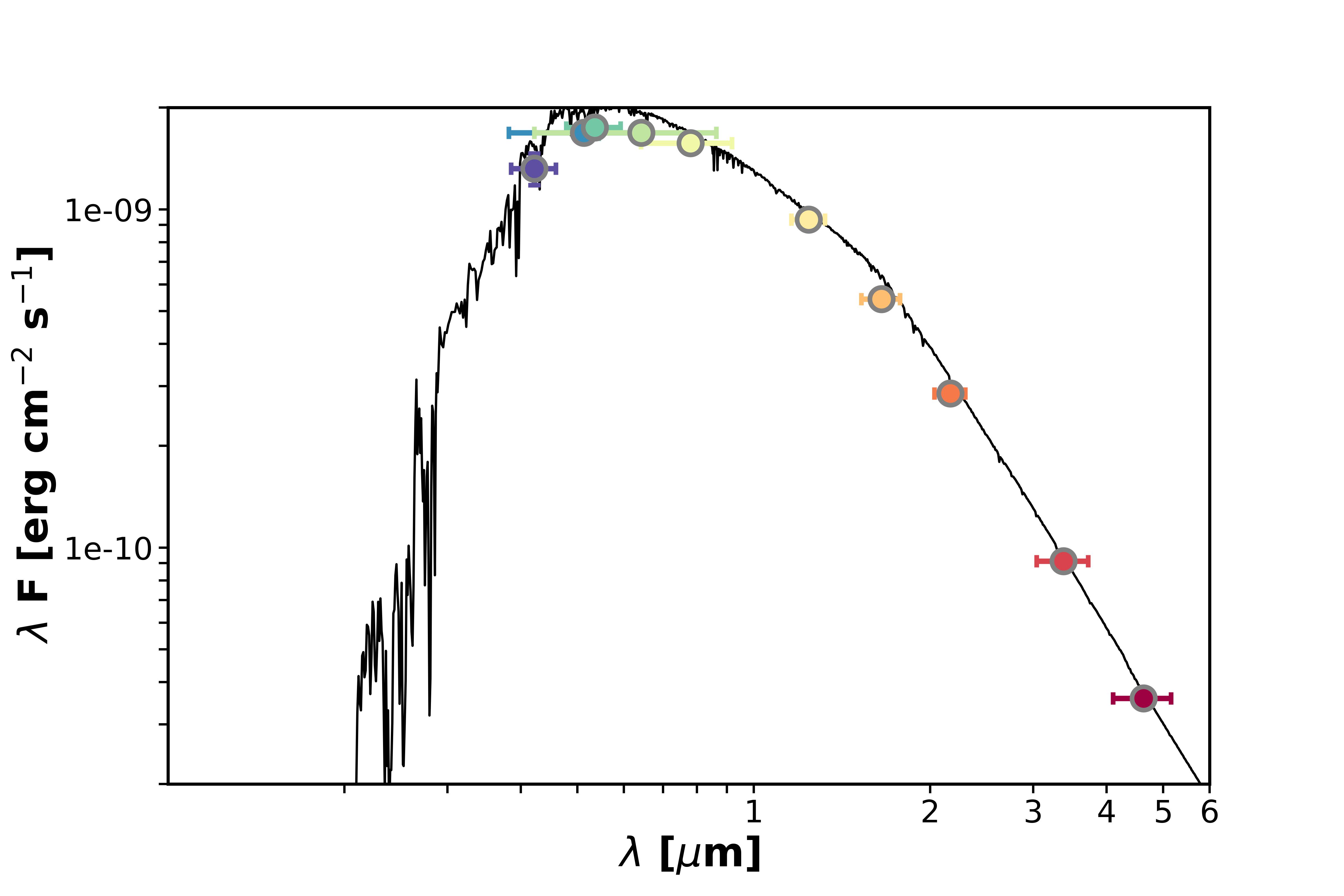}\hfill
\caption{Spectral energy distribution (SED) of \target. Circles indicate the photometric bands with horizontal bars reflecting the passband width.}
\label{fig:SED}
\end{figure}

\begin{figure}
\includegraphics[width=.49\textwidth]{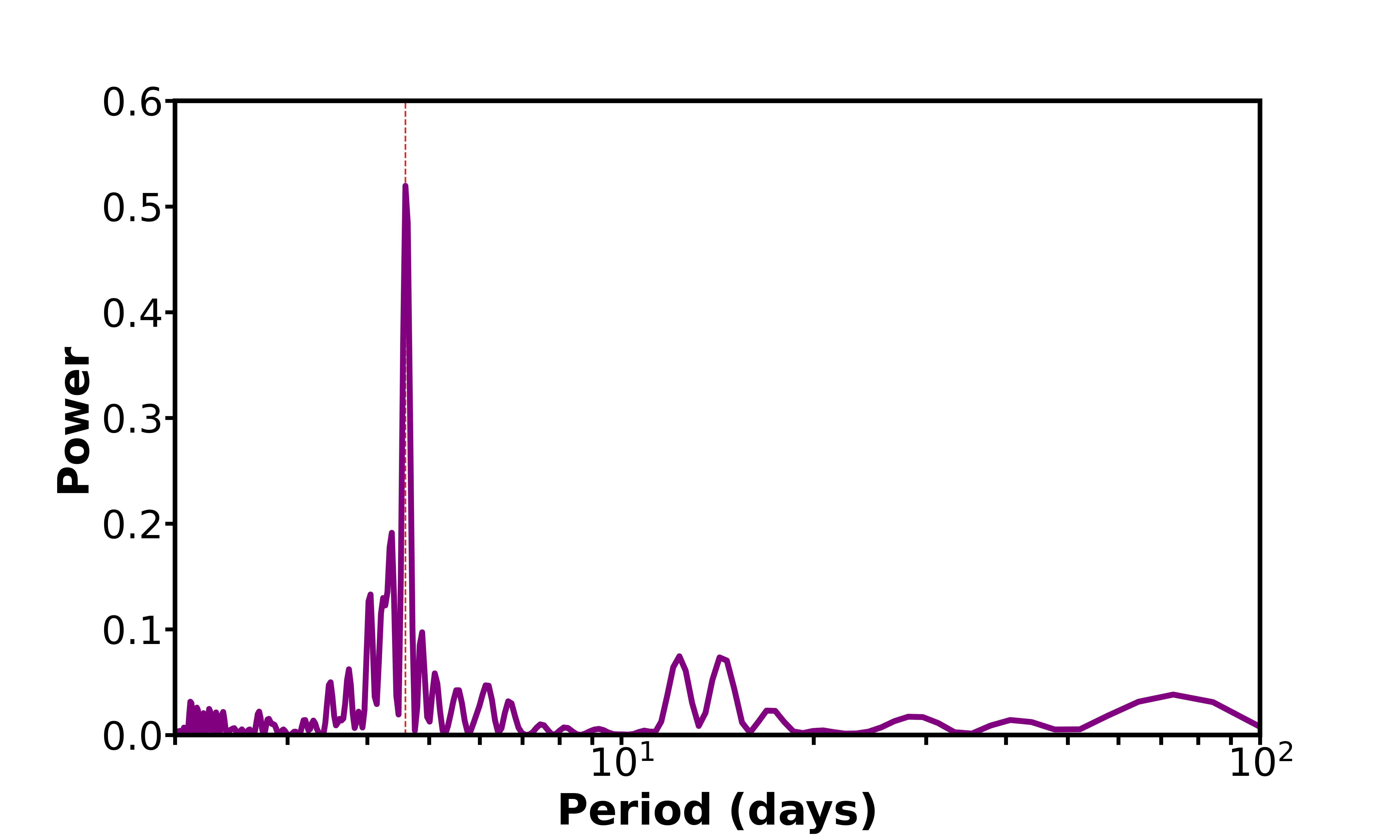}\hfill
\caption{Lomb-Scargle periodogram of the raw SPOC \target{} \lc. The highest-power period is denoted by a dashed line. Uncertainties are derived from FWHM of the highest-power resulting in a period of \Prot{} days.}
\label{fig:periodogram}
\end{figure}

\subsection{Global Modeling}
\label{ssec:globalfit}

To characterise the physical parameters of the \target{} system, along with their associated uncertainties, we performed a Markov Chain Monte Carlo (MCMC) fit using the publicily available software package \textsc{emcee} \citep{emcee}. We opted to not include the 2022-Apr-01 CTIO observation of \bPlanet{} in our fit because it only captured an ingress. Detrending was performed simultaneously to the transit fits for each photometric observation. During each iteration, we subtracted a transit model created from the \textsc{batman} package \citep{batman} from the observed data. Next, we fit a trend model to the residuals using least-squares optimization, and compared the best trend model and the residuals to determine the log likelihood for the sampler. Section \ref{sec:data} contains further information on the individual trend models.

The global modeling process involved fitting for the two planets simultaneously. Our fit included the period, epoch, the planet-to-star radius ratio, {\rpl}/{\rstar} and  impact parameter, {$b$} of both planets. Eccentricities were initially fixed at 0 as we assumed a circular solution. The limb darkening coefficients for each unique photometric filter, denoted as $q_{1,i}$ and $q_{2,i}$, were included in the global fit. These coefficients were parameterised as per \citet{kipping2013}. Additionally, we included a scaling factor for each observation as a free parameter in the fit, {$c_i$}. This accounts for differing levels of background contamination due to the varying apertures of the individual instruments. This parameter is neccessary as the large PSFs of both \tess{} and \cheops{} allow background stars to contribute to the measured flux, resulting in differing total fluxes to that of the LCOGT \lcs. We also opted to fit for \mstar{} and \rstar{} to simultaneously constrain them with the planet parameters. 

To constrain the period and transit times of both planets, we imposed uniform priors with boundaries at the 5-$\sigma$ upper and lower uncertainty provided by the SPOC detection pipeline. For \tess{} and \cheops, we allow the $c$ values to uniformly explore between boundaries of 0.9 to 1.1, this is due to the brightest contaminating star in TESS/CHEOPS aperture account for roughtly 5\% of the light of the target star. Since the LCOGT aperture excluded all neighboring stars resolved by the Gaia catalog, we tightly constrain $c_{lco}$ to only uniformly vary between 0.99 and 1.01. For \mstar{} and \rstar, we imposed a Gaussian prior derived from the results of the SED fitting obtained in Section \ref{ssec:stellarparameters}.  

We first assumed the orbits of the two planets to be circular. We also ran a set of global modeling allowing non-zero eccentricities for both planets. Our eccentric fit consisted of all previously mentioned free parameters, as well as the addition of \ecosw{} and \esinw{} as free parameters, where $e$ is the eccentricity and $\omega$ is the longitude of periastron. These parameters could be derived to find the eccentricity and longitude of periastron of both planets. The results of both fits can be seen in Table \ref{tab:planettable}. 

Through visual examination of the detrended \lcs{} against the models, assuming a constant period, we noticed that both planets deviated slightly from their predicted transit timing assuming they are both on circular orbits. We followed the algorithm outlined in Nabbie et al. (in prep) to search for potential transit timing variations (TTVs). The 2022-Apr-01 CTIO observation of \bPlanet{} and Teid observation of \cPlanet{} were excluded due to their insufficient out-of-transit baselines. For the rest of the transits, we fit for their individual transit centers independently while allowing for the trend parameters to vary similarly to the previous model. Instead of using Gaussian priors on the stellar parameters, we added the likelihood term from the SED in our model instead so that all the stellar parameters can be constrained at the same time as the transit parameters. 

For the SED likelihood, we make use of the isochrone package \citep{2015ascl.soft03010M} and the MIST tracks. We fixed stellar parallax, distance and reddening to the best-fit values from \ref{ssec:stellarparameters}. We added {\teff}, {\feh} and age as our free parameters, and an error scale parameter following \citet{Eastman:2019}. The priors of {\teff} and {\feh} are constrained by Gaussian priors with uncertainties provided by Section \ref{ssec:stellarparameters}. We adopt uniform priors for {\mstar} and {\rstar} with bounds between 1 and 1.4. The error scale was bounded between 0.01 and 100, and stellar age between 0.5 and 10 Gyr. 

\section{Results and Discussion}
\label{sec:discussion}
\label{section:discussion}

\subsection{Global modeling results}

Through the results of our analysis conducted in Section \ref{ssec:globalfit}, we report the discovery and validation of two planets around the bright F star \target. We have determined \bPlanet{} to be a hot super-Neptune, with a radius of \bRadius \rearth{} in a \bPeriod{} day orbit. Further, we have discovered and validated \cPlanet{} as a warm Neptune at \cRadius \rearth. \cPlanet{} orbits its host star every \cPeriod{} days. Comparison with the current exoplanet demographic is found in Figure \ref{fig:irradiance_pop_plot}. The full parameter results are presented in Table \ref{tab:planettable}. The eccentric fit and the circular fit have predominantly overlapped posterior space. We have also detected a tentative TTV signal which will be discussed in Section \S \ref{sssec:ttv_lc}. As expected, the best fit parameters retrieved when allowing the transit times to vary in the global model, produced slightly larger transit depths and shorter transit durations for both planets. However, the joint modeling with the SED lead to a slightly smaller stellar radius. The posterior of the final planet parameters largely overlap, with the TTV fit reports \bPlanet{} having a radius of \bRadiusttv \rearth{} and a transit duration of  \bDurationttv{} hours. For \cPlanet{}, the TTV fit reports a radius of \cRadiusttv \rearth{} and a transit duration of  \cDurationttv{} hours. 

\subsection{Validation of the TOI-5126 system}

Both astrophysical phenomena and instrumental artifacts can replicate the signal of a planetary transit. Without careful consideration of false positive scenarios, the legitimacy of the planet can be compromised. In this section, we validate the \target{} system by ruling out false positive scenarios which could have created the observed transit signals. We consider the following scenarios:
\begin{itemize}
    \item \textbf{Either of the transit signals are a reflection of instrumental artifact:} Both transit signals were detected with high significance in the \tess{} data, \cheops{} and ground-based telescopes as mentioned in section \ref{sssec:groundbasedphotometry}, therefore this possibility is ruled out.

    \item \textbf{\target{} is an Eclipsing Binary}: This scenario is ruled out by our radial velocity observations from TRES and CHIRON (section \S\ref{ssec:tres}). An analysis on the system shows that a brown dwarf companion (\mpl $\approx$ 13\mjup{}) orbiting around \target{} would produce RV semi-amplitudes of \Kb = \Kbbrown{} \ms{} and \Kc = \Kcbrown{} \ms. Neither sets of the radial velocity observations show such deviations.

    \item \textbf{Light from a distant eclipsing binary or transiting planet system is blended with \target}: The magnitude difference between \target{} and the faintest companion which could replicate the signals observed follow as: $ \Delta m \lesssim 2.5 \log10 \left( \frac{\ingress ^2}{\tduronethree ^2 \delta_{obs}} \right) $. The faintest star which could replicate the transit signals for \bPlanet{} is $\Delta$$m$ $=$ 7, and \cPlanet{} is $\Delta$$m$ $=$ 6. Direct imaging observations rule out stars with a $\Delta$$m$ of 7 from 0.5\arcsec{} onwards, and ground-based photometric observations detected the transit events within a  4.7\arcsec{} aperture. To statistically rule out stars within 0.5\arcsec{} of \target, we determined the average stellar density in the nearby field from Gaia data. For stars within a magnitude range of seven magnitudes, the average stellar density is calculated to be {$2.4\times10^{-5}$ arcsec$^{-2}$}. Narrowing this caclulation to stars six magnitudes fainter than \target, the average stellar density is {$4.0\times 10^{-5}$ arcsec$^{-2}$}. 
\end{itemize}

An additional investigation into the probability of remaining false positive scenarios was performed using \textsc{Triceratops} \citep{triceratops}. We used the detrended \tess{} \lcs{} of \bPlanet{} and \cPlanet{}, as well as constraints provided by Palomar imaging (see section \ref{ssec:imaging}) in our analysis. A false-positive probability (FPP) of 0.77\% and a nearby false-positive probability (NFPP) of 0\% offer statistical validation for \bPlanet. Analysis of the \tess{} \lc{} for planet c reveals a FPP and NFPP of 5.33\% and 4.98\%, respectively. This is invoked by a Nearby Eclipsing Binary scenario on 9.53\arcsec{} away neighbour TIC27064467. This scenario, however, has been ruled out by the on-target event detected by ground-based photometry. As a result, the FPP with this scenario eliminated is 0.35\%.

\subsection{Dynamical constraints on the system}
\label{ssec:dynamics}

\subsubsection{Long-term stability}
\label{sec:long-term stability}

The masses and eccentricities of the planets are currently not well-constrained. In this section, we explore the possibility of placing additional constraints on the system through considerations of long-term dynamical stability. We first assess the stability of the observed system using calculations of the Mean Exponential Growth factor of Nearby Orbits (MEGNO; \citealt{2000A&AS..147..205C}; \citealt{2003PhyD..182..151C}) indicator within the REBOUND gravitational dynamics software package \citep{2012A&A...537A.128R}. For a given set of initial conditions, we quantify the stability of the system using the time-averaged MEGNO chaos indicator value, $\langle Y(t) \rangle$, which reflects the degree of divergence of initially closely-separated trajectories in phase space. Stable system configurations will have MEGNO values $\lesssim$ 2. 

The MEGNO value across a grid of different masses for both planets is displayed in the top panel of Figure 8. We are able to obtain a relatively loose constraint on the planet masses: approximately, if $M_b+\frac{2}{3}\,M_c<3 M_J$, the system will be stable. To estimate the average MEGNO value in each cell, we randomly sample the initial condition 30 times. We assume eccentricities of both planets were randomly drawn from a Rayleigh distribution with scale parameter $0.1$. The mean anomaly and argument of periapse angles are sampled uniformly between 0 and 2 $\pi$. The integration time for each realization was set to $10^6$ orbits of the outermost planet with a time-step of $1/20$th of the innermost planet's period. 

We also found the system will be stable across a wide range of eccentricities for both planets if their masses approximated by the mass-radius relationship following \cite{Otegi:2020}. Therefore, the stability criterion alone cannot provide additional constraint on the system eccentricity. 

\begin{figure}
\includegraphics[width=.45\textwidth]{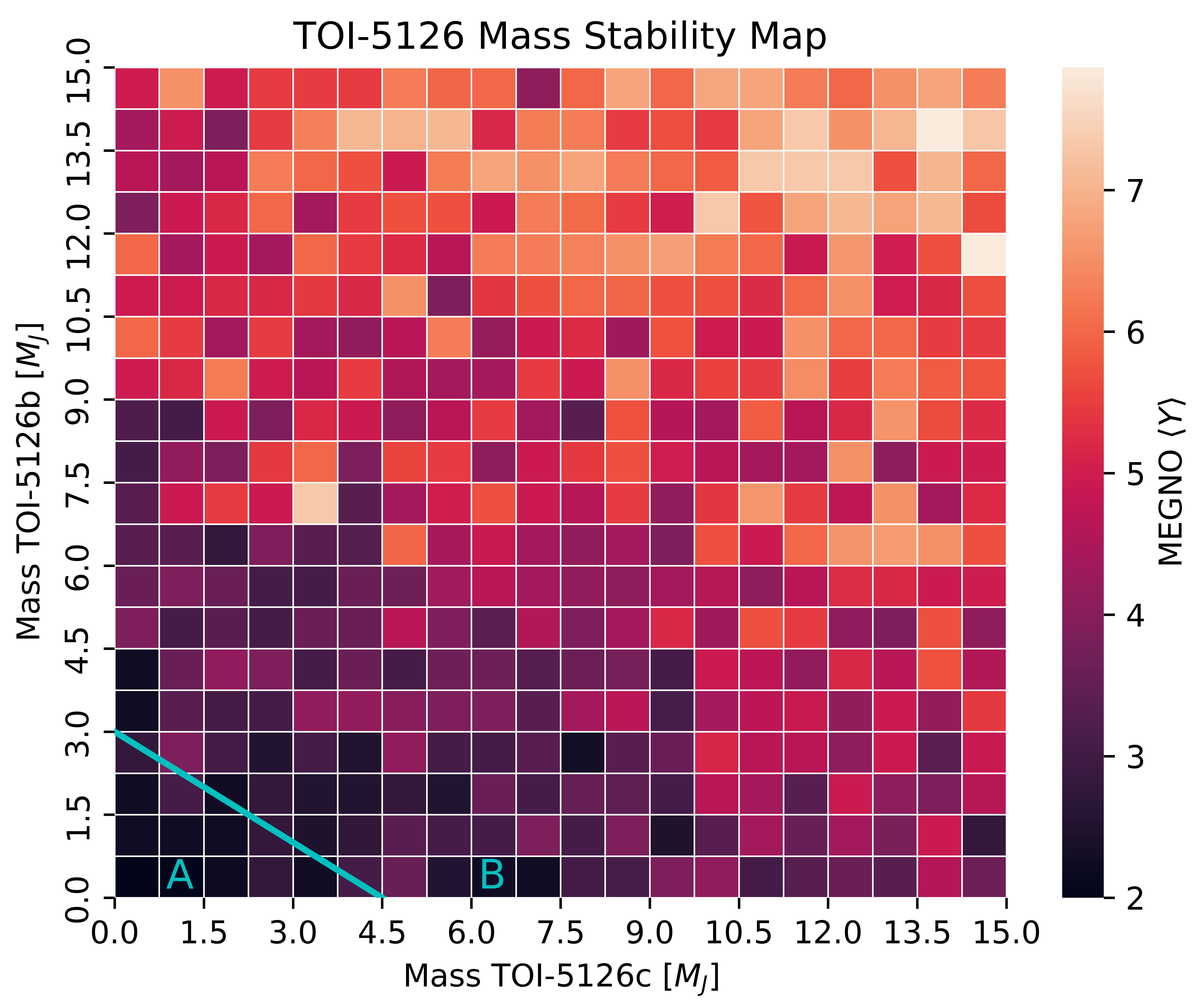}
\includegraphics[width=.45\textwidth]{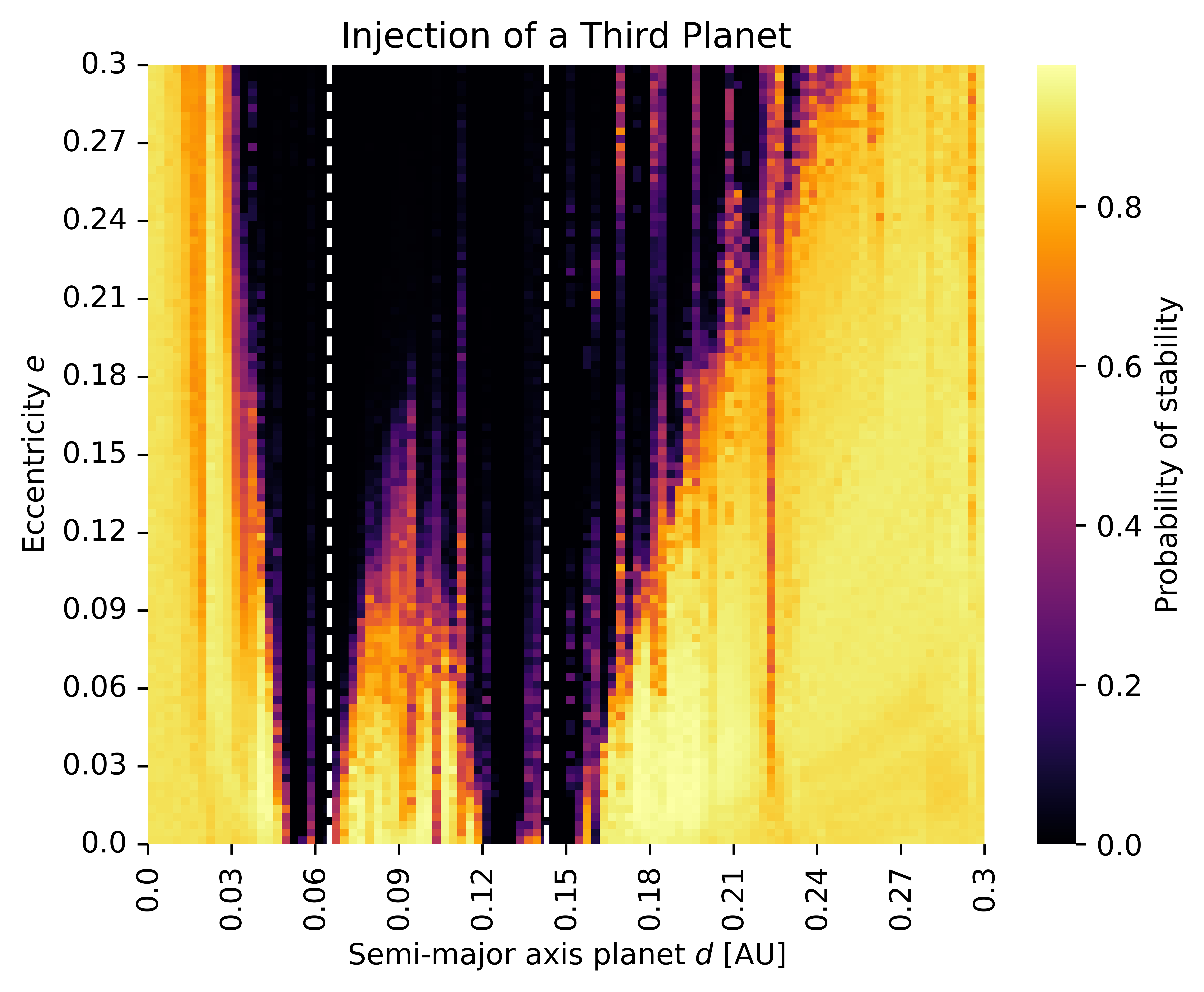}
\caption{\textit{Top}: MEGNO stability map in the $M_b - M_c$ parameter space. Two distinct regions of stability are divided by the solid cyan line, with the region A below the line indicate stable configurations. Each pixel in the $20 \times 20$ map holds the average of $30$ realizations of randomized initial conditions. \\
\textit{Bottom}: Stability map generated using SPOCK for a hypothetical third planet within the \target{} system. The vertical white dashed lines denote the locations of \bPlanet{} and \cPlanet. Each cell in the map represents the average of 10 realizations with randomized initial conditions. There is a small stable region present between the two planets in which a hypothetical third planet could reside.}
\label{fig:stability results}
\end{figure}

\subsubsection{Stability of a hypothetical third planet}

The two detected planets in the \target{} system are fairly widely-spaced ($P_c/P_b \sim 3.3$). It is possible that there is a third planet residing between the orbits of the two planets or somewhere else in the system. To explore this possibility, we use the Stability of Planetary Orbital Configurations Klassifier (SPOCK; \citealt{2020PNAS..11718194T}) to compute the probability of stability of a given set of parameters. SPOCK is a machine learning model trained on $\sim 100,000$ orbital configurations of three-planet systems. It can estimate the probability of long-term stability of planetary systems with three or more planets.

In particular, we explored the $a_d - e_d$ space of the additional planet, whose mass we set to the average of the nominal masses of planets b and c based on the mass-radius relationship discussed in Section \ref{sec:long-term stability}. We assumed circular orbits for planets b and c while varying the eccentricity of planet d from $0 - 0.3$ and semi-major axis from $0 - 0.3$ AU. Setting the orbits of the known planets to circular provides us with the maximum possible stable region for the hypothetical third planet. Each grid cell in the $a_d - e_d$ map represents the average of 10 realizations of the system with randomized initial conditions. Specifically, the mean anomalies, longitudes of ascending nodes, and inclinations were randomly drawn according to $M, \omega \sim \mathrm{Unif}[0, 2\pi]$, $i \sim \mathrm{Rayleigh}(1^{\circ})$. 

The bottom panel of Figure \ref{fig:stability results} shows the results of the SPOCK stability map for a hypothetical third planet. In addition to stable regions interior to \bPlanet{} and exterior to \cPlanet{}, there is also a stable region between the two planets. It is possible that a planet may be transiting but remain undetected due to its small radius. Alternatively, the planet could be also be either non-transiting or in the trivial case not exist at all. Nevertheless, Figure \ref{fig:stability results} indicates the ideal orbital proximity in which to search for additional planets in the system.

\subsubsection{Transit timing variations signal in the light curves}
\label{sssec:ttv_lc}

\begin{figure*}
\centering
\includegraphics[width=.8\textwidth]{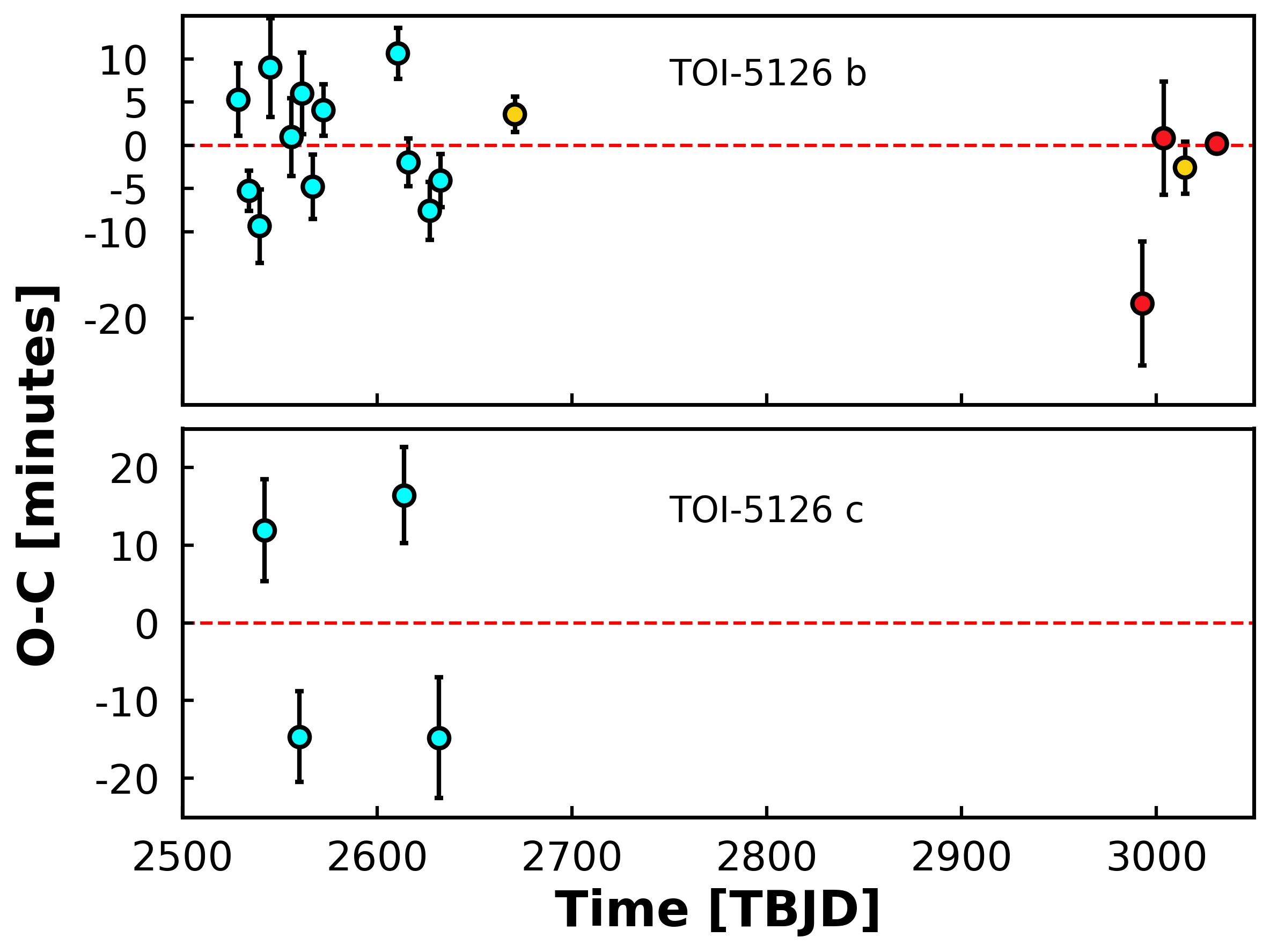}
\caption{Transit times subtracted from a linear ephemeris for \bPlanet{} and \cPlanet. Cyan indicates \tess{} transits, yellow indicates LCOGT transits and red indicates \cheops{} transits. The red dashed line shows the expected transit time assuming a constant period. }
\label{fig:ttvplot}
\end{figure*}

From our analysis of the {\tess} \lcs, we find that both planets exhibit tentative TTVs, seen in Figure \ref{fig:ttvplot}. Such signals can potentially arise due to the gravitational perturbations that the planets impart on one another as they reach conjunction, creating non-periodic transit times. Notably, {\bPlanet} and {\cPlanet} are not near a first-order mean-motion resonance (MMR), which would produce the largest TTV amplitudes \citep{agol2005}. 

We utilised REBOUND \citep{2012A&A...537A.128R} to estimate expected TTV signals based on \bPlanet{} and \cPlanet{} alone. We sample the initial orbital elements of both planets ($a$, $e$, $w$, $i$) from the posterior distribution results from our global models (including the TTV fit) in Section \ref{sec:analysis}. The other orbit angles are sampled uniformly between 0 and 2\,$\pi$. The systems are integrated for more than 500 days and transit times corresponding to the observed transits from \tess, LCO and \cheops{} are recorded. The expected TTV scatter is then computed by calculating the standard deviation of the transit times after subtracting the best fit linear ephemeris. We simulated the system for 10000 realisations and found that \bPlanet{} produced 2-$\sigma$ upper limit variations of \bttvtest{} minutes, with \cPlanet{} exhibiting \cttvtest{} minute 2-$\sigma$ upper limit TTVs. The current system is therefore unable to explain the observed TTVs and we consider alternate scenarios.

One possible interpretation is that these signals imply the existence of a third, non-transiting planet in between \bPlanet{} and \cPlanet{} (as discussed in the previous section), so that one of the neighboring planet pairs have a period ratio just wide of a 3:2 resonance and the other just wide of a 2:1 resonance, both of these period ratios being commonly found in multi-planet systems \citep{Fabrycky:2014}.

However, we caution that it has been shown in previous studies that spot crossing events can mimic TTV signals rather than gravitational interactions due to planets \citep{Sanchis-Ojeda:2011, Fabrycky:2012, Szabo:2013, Mazeh:2013}, especially when the individual transit has relatively low signal-to-noise. \citet{Ioannidis:2016} demonstrated that a SNR larger than 15 is required to statistically minimizes the probability that a spot crossing event being the cause of the deviations. The SNR of planet b in \tess{} data is 13, with planet c having a SNR of 11. Given two out of three of our CHEOPS observations does not cover either ingress or egress of the transits, further high signal-to-noise photometry follow up is needed to unlock the full system architecture of \target.

\subsection{\bPlanet{} in context with the Super-Neptune Population}

\bPlanet{} is the newest addition to the super-Neptune population, the class of planets with radii between 4 and 8 \rearth{}, and diverse compositions (with mass ranging between  6-135 \mearth{} \citealt{NEA}). \bPlanet{}'s planetary parameters 
(\rpb $\approx$ \bRadiusapprox{} \rearth, $P_{b} \approx$ 
\bPeriodapprox{} days) 
place it in the sparsely populated $\lesssim$10 day, 4-8 \rearth{} region of the period-radius plane, known as the `hot Neptune desert' \citep{Mazeh:2016} (see Figure \ref{fig:irradiance_pop_plot}) 

\begin{figure}
\centering
\includegraphics[width=.49\textwidth]
{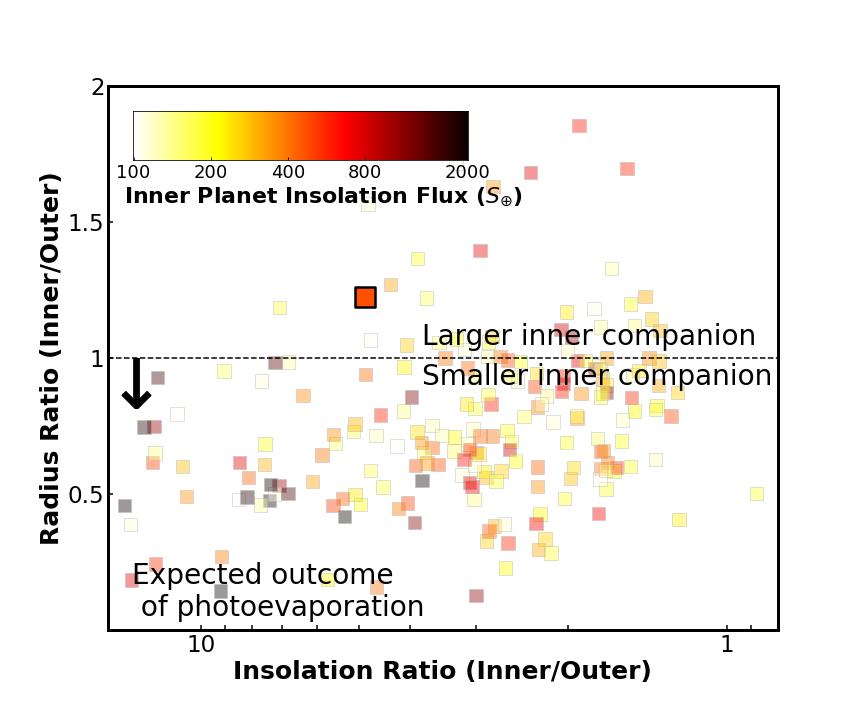}\hfill
\caption{Radius ratios of all known inner/outer planet pairs as a function of their insolation ratios. The colours represent the inner planets' insolation flux in Earth units. The horizontal dashed line at one represents a same sized planet pair, with the arrow outlining the decrease in radius ratio expected by photoevaporation of the inner hotter planet. This plot makes use of data from the NASA exoplanet archive \citep{NEA} downloaded on 2023-02-07 UT. The data extracted only planets with measured periods, radii and host star masses, radii and effective temperatures known.}
\label{fig:ratioplot}
\end{figure}

\begin{figure}
\centering
\includegraphics[width=.49\textwidth]{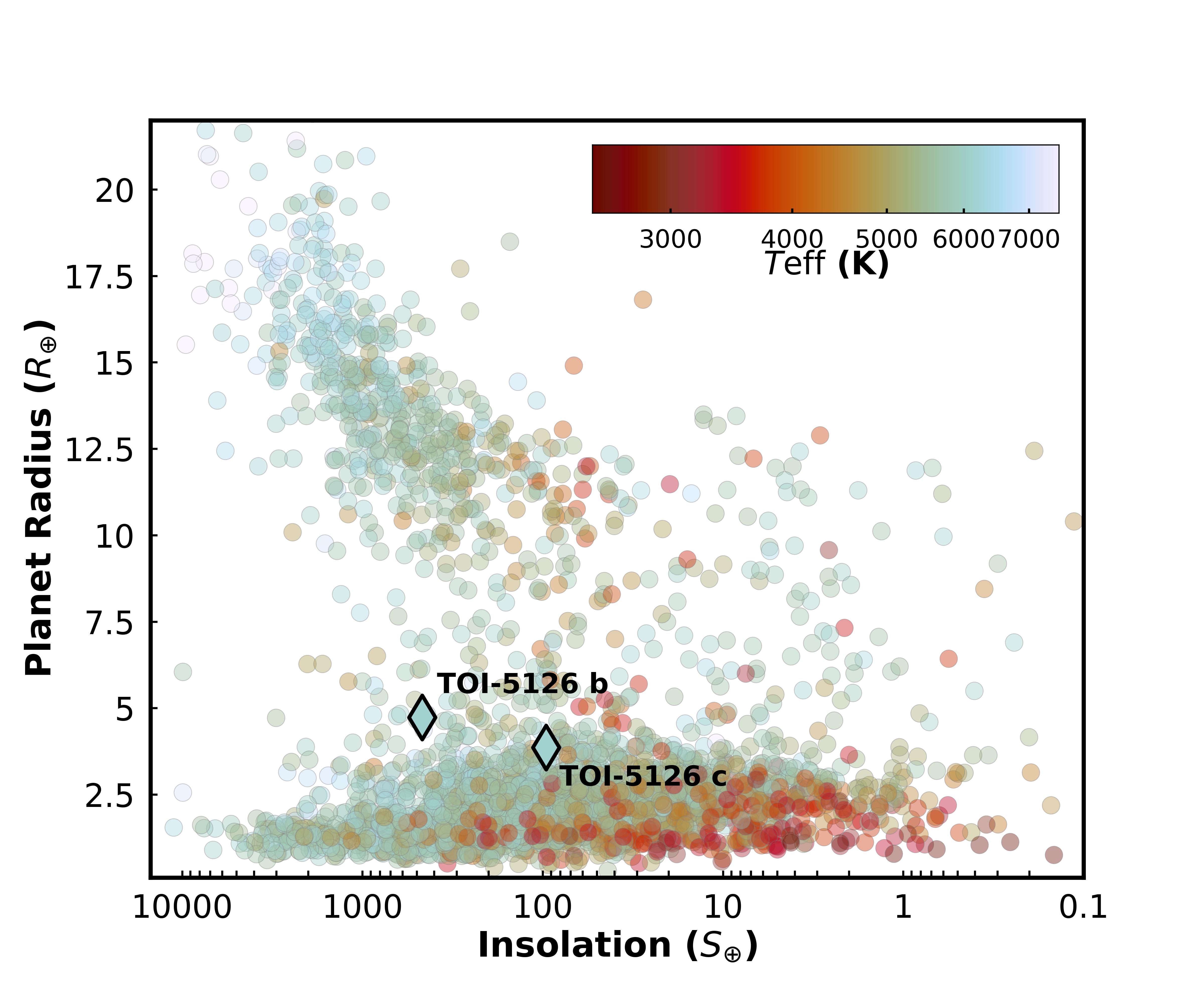}\hfill
\caption{Radius relative to insolation flux for the population of confirmed planets. Colours represent the host star effective temperature. Plots were made with data downloaded from NASA exoplanet archive on 2023-02-07 UT.}
\label{fig:irradiance_pop_plot}
\end{figure}

The unusually large radius of \bPlanet{} implies that long-term mass-loss processes are likely still ongoing and may be readily observable. Of all super-Neptunes in the desert, \bPlanet{} is on an exclusive list of three (WASP-166\,b, \citealt{Hellier:2019}; LTT 9779b, \citealt{Jenkins:2020}) with visual magnitudes $\lesssim$10 and predicted equilibrium temperatures above the cloud-free threshold ($\gtrsim$1200 K). This makes it one of the most readily characterisable highly irradiated super-Neptune's for transmission spectroscopic observations, where mass-loss mechanisms can be directly probed. Questions surrounding mass-loss are particularly unique for \target{} as planet b is larger than planet c, despite receiving more stellar irradiation. This is in contrast to the expected outcome of photoevaporation, which implies that the inner companion would be smaller if the insolation ratio is larger. An example of the radius and insolation ratio relation, as well as the expected outcome of photoevaporation, can be seen in Figure \ref{fig:ratioplot}. 

\subsection{Other future follow up opportunities}
\label{ssec:followup}

We compute predicted masses for \target{} b and c based on the mass-radius relationship from \citet{Otegi:2020}. These predicted masses are adopted for the discussions below. The predicted masses for \target{} b and c are $\bMassOtegi{}$ \mearth{} and $\cMassOtegi{}$ \mearth, respectively.

The estimated RV semi-amplitudes for planets b and c are \Kb{} $=$ \bPredictedK{} \ms{} and \Kc{} $=$ \cPredictedK{} \ms{}, respectively. Measuring the masses of both planets are important to determine if they are formed with large protoplanetary cores and therefore a large gaseous envelope \citep{LeeChiang2016} or inflated atmosphere due to tidal heating \citep{Millholland2020}. \target{}'s high rotational velocity makes RV measurements difficult, but with the next generation of instruments such as MAROON-X \citep{MAROONX}, along with innovative methods to correct for stellar variability, mass constraints of both planets are not ruled out. The rapid rotation of \target{} does however enable follow-up studies into the spin-orbit alignment of the system with the Rossiter-McLaughlin effect (R-M, \citealt{rossiter1924, mclaughlin1924}). \bPlanet{} has a predicted R-M semi-amplitude \Kb{} $=$ \bRM{} \ms, and planet c \Kc{} $=$ \cRM{} \ms. 

The planets in the \target{} system present an ideal opportunity to better understand the atmospheric composition difference between a hot super-Neptune and a warm Neptune. \bPlanet{} still likely retains a significant H/He atmosphere due to its inflated radius. 

\begin{figure}
\centering
\includegraphics[width=.49\textwidth]{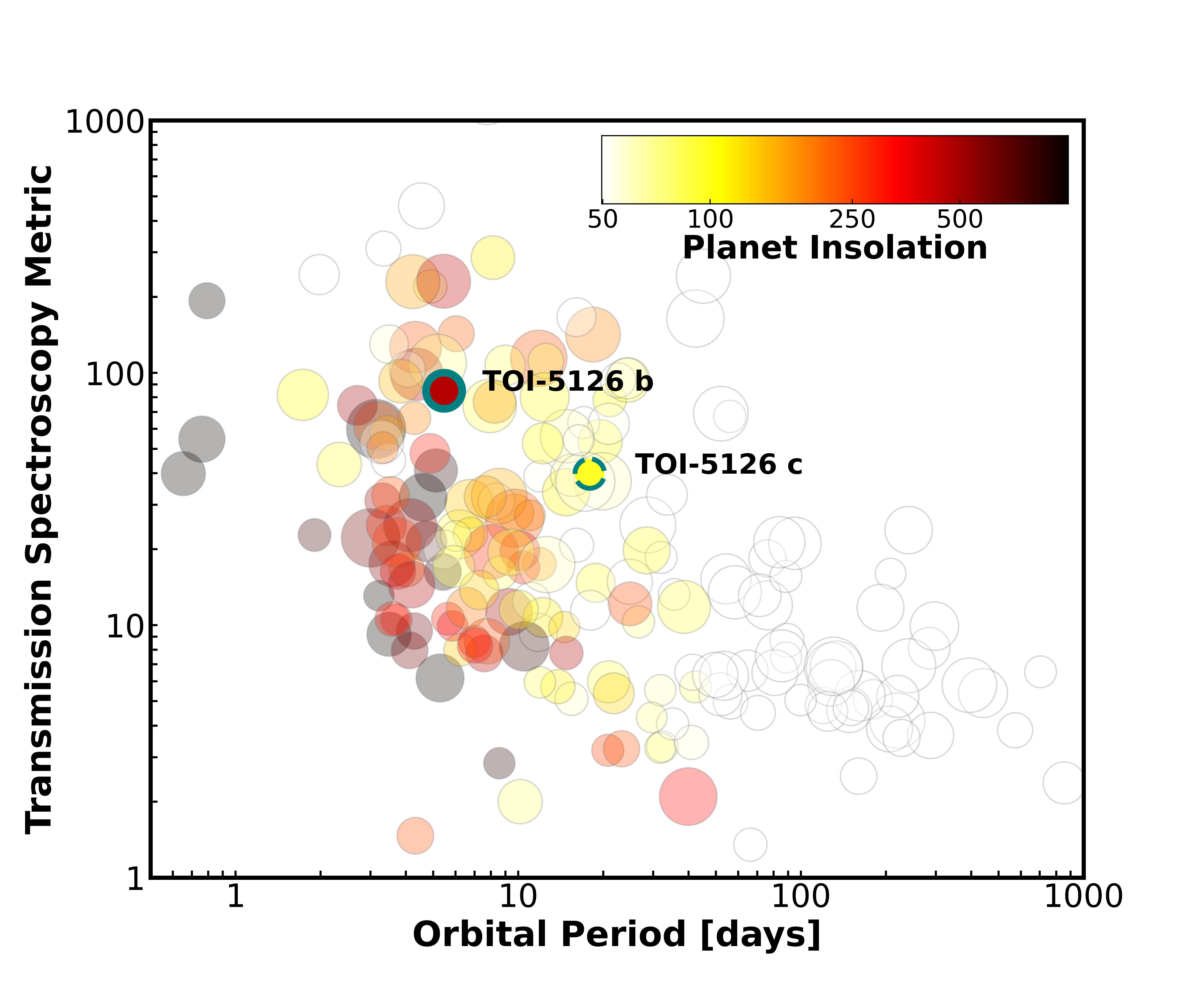}\hfill
\caption{\target{} planets among all super-Neptunes in terms of orbital period and Transmission Spectroscopy Metric (TSM). \cPlanet{} is notably not a super-Neptune and is outlined for comparative purposes as a dashed line. The color of each point describes the planets' insolation flux, with point size representing their scaled radius. This plot makes use of data from the NASA exoplanet archive downloaded on 2023-02-07 UT.} 
\label{fig:tsmcontextplot}
\end{figure}

\target{} is relatively bright, with a $J$ band magnitude of 9.04. The inner planet \bPlanet{} has an equilibrium temperature of (\teq $=$ \bTeq{} K), and should host a relatively clear atmosphere. To test its suitability for follow-up, we followed the transmission spectroscopy metric (TSM; \citealt{kempton2018}) and find that \bPlanet{} has a TSM of \bTSM. Given the potential of \cPlanet{} to also be a candidate for transmission spectroscopy (TSM $=$ \cTSM), this system enables the potential of atmospheric comparative planetology studies between a hot super-Neptune and warm Neptune. The suitability of the \target{} planets for follow-up studies can be seen in comparison to the super-Neptune population in Figure \ref{fig:tsmcontextplot}, where despite \cPlanet{} being a Neptune, its high insolation flux (\Insolc{} = \cIrradiation{} \Searth) and predicted equilibrium temperature (\Teq{} = \cTeq{} K) make it favourable even among super-Neptunes. 

\begin{figure*}
\centering
\includegraphics[width=.99\textwidth]{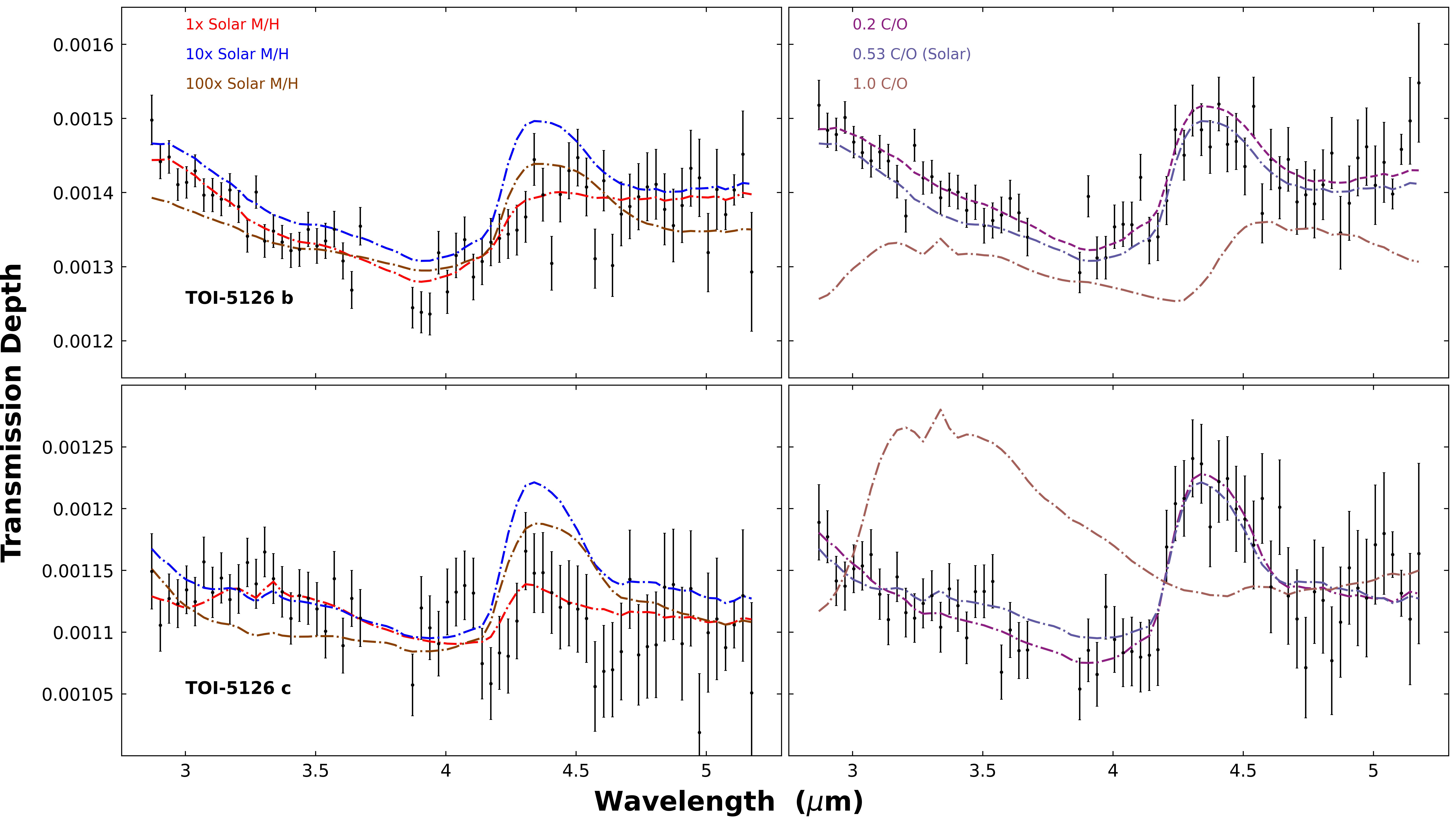}
\caption{Simulated JWST Transmission Spectrums of both planets with varying metallicities and C/O ratios. \textbf{Left}: The red, blue and brown lines represent the varied models of 1x, 10x and 100x Solar M/H respectively. \textbf{Right}: C/O simulations were based off a 10x M/H baseline. the purple, light blue and brown lines indicate 0.2, 0.53 (Solar) and 1.0 C/O ratios.}
\label{fig:jwstsim}
\end{figure*}

To demonstrate the possibility of comparative studies, we conducted preliminary atmospheric simulations using the public software package \textsc{petitRADTRANS} \citep{petitradtrans}. As we are focused on identifying differences in the M/H and C/O ratios of the two planets, we used the NIRSpec Bright Object Time Series (BOTS) template with a G395H grating for our simulations. We first simulated varying M/H, including 1x, 10x and 100x Solar. We also simulate the different C/O ratios at 0.2, 0.53 (Solar) and 1.0 C/O levels, in this case assuming 10x solar M/H. The results can be seen in Figure \ref{fig:jwstsim}.

The key features of the G395H grating are the {$\mathrm{CH_4}$} and {$\mathrm{CO_2}$} features in the spectra where both planets can be probed to compare formation theories. In this case, a high C/O ratio and metallicity would imply formation in a gas-depleted disk, while low C/O ratios for the planets would support other formation mechanisms. We show that differing {$\mathrm{CO_2}$} abundances can be discerned in both \bPlanet{} and \cPlanet{} in the M/H simulation, while \cPlanet{} appears to also have a small {$\mathrm{CH_4}$} feature at lower metallicities due to its lower predicted equilibrium temperature. The varying C/O models show that if \cPlanet{} has a high  {$\mathrm{CH_4}$} abundance, its chemical signature will be apparent in the spectrum, while lower levels would as a result produce a larger {$\mathrm{CO_2}$} bump. The high equilibrium temperature of \bPlanet{} diminishes the signature of the {$\mathrm{CH_4}$} feature at high C/O ratios. 

\section*{Acknowledgements}
\label{sec:acknowledgements}
%

We respectfully acknowledge the traditional custodians of the lands on which we conducted this research and throughout Australia. We recognize their continued cultural and spiritual connection to the land, waterways, cosmos and community. We pay our deepest respects to all Elders, present and emerging people of the Giabal, Jarowair and Kambuwal nations, upon whose lands the MINERVA-Australis facility at Mount Kent is located.

TF's research is funded by the University of Southern Queensland Undergraduate Research Scholarship Program. EN acknowledge the PhD scholarship provided by the ARC discovery grant DP220100365. CH thanks the support of the ARC DECRA program DE200101840. GZ thanks the support of the ARC DECRA program DE210101893. 

DRC acknowledges partial support from NASA Grant 18-2XRP18\_2-0007. 

CHEOPS is an ESA mission in partnership with Switzerland with important contributions to the payload and the ground segment from Austria, Belgium, France, Germany, Hungary, Italy, Portugal, Spain, Sweden and the United Kingdom. We thank support from the CHEOPS GO Programme and Science Operations Centre for help in the preparation and analysis of the CHEOPS observations.

This work has made use of data from the European Space Agency (ESA) mission Gaia (https://www.cosmos.esa.int/gaia), processed by the Gaia Data Processing and Analysis Consortium (DPAC, https://www.cosmos.esa.int/web/gaia/dpac/consortium).

B.S.S. and A.A.B acknowledge the support of M.V. Lomonosov Moscow State University Program of Development.

Some of the observations in the paper made use of the NN-EXPLORE Exoplanet and Stellar Speckle Imager (NESSI). NESSI was funded by the NASA Exoplanet Exploration Program and the NASA Ames Research Center. NESSI was built at the Ames Research Center by Steve B. Howell, Nic Scott, Elliott P. Horch, and Emmett Quigley. Data were reduced using a software pipeline originally written by Elliott Horch and Mark Everett.

This work makes use of observations from the LCOGT network. Part of the LCOGT telescope time was granted by NOIRLab through the Mid-Scale Innovations Program (MSIP). MSIP is funded by NSF. 

This research has used data from the CTIO/SMARTS 1.5m telescope, which is
operated as part of the SMARTS Consortium by \href{http://secure-web.cisco.com/1TL5nionOJJUGi7T0X_YvX7RLRwbVQl20QG7s4LKeK1vpFY8M3UHYMuONVvV2D2hxli_pMi4YkHdTYel4ogZ3sJWN4axM8-5IsyCIPeIj7BfVIBOvp9a8iRKv2IM-wTBpjGA3xxZcH5lT4FNKBIoEstyJEEyUYzEKbDQyL4T1LQSiukl5eTarVlkS9YJbHf_HrjiuXV1gM1uXr7gdIdCbZg4CfJa_N8Qw38oz0KhpJ74RZ0rIcyg3XKCc6-HCDjlBrMtX3cpMKa1Kcya1SxY0UxXY0WkwM0zGeXYUYfbkp1Ce6jIBY8Evcz-YcyODRE4QWMlPqSDV66bKv5F1R3-RrkcH91Y7INyFOP6qJfGJKLRFJT-KNphpqmNc4Pf7zLVOIBjCEKsANmt1XTtzQN5AIPwKf-F1qd4b6KCZrqjHZIA/http\%3A\%2F\%2Fwww.recons.org}{RECONS}

This research has made use of the NASA Exoplanet Archive, which is operated by the California Institute of Technology, under contract with the National Aeronautics and Space Administration under the Exoplanet Exploration Program. 


This research has made use of the Exoplanet Follow-up Observation Program (ExoFOP; DOI: 10.26134/ExoFOP5) website, which is operated by the California Institute of Technology, under contract with the National Aeronautics and Space Administration under the Exoplanet Exploration Program.

We acknowledge the use of public TESS data from pipelines at the TESS Science Office and at the TESS Science Processing Operations Center. 

Resources supporting this work were provided by the NASA High-End Computing (HEC) Program through the NASA Advanced Supercomputing (NAS) Division at Ames Research Center for the production of the SPOC data products.

Funding for the TESS mission is provided by NASA's Science Mission Directorate. KAC acknowledges support from the TESS mission via subaward s3449 from MIT.

\textit{Facility}: TESS, CHEOPS, Exoplanet Archive, McD 1.0 m, Teid 1.0 m, CTIO
1.0 m, CTIO/CHIRON 1.5 m, LCOGT, Palomar/PHARO 5.0 m, WIYN/NESSI 3.5 m, SAI 2.5 m, FLWO/TRES 1.5 m.\\

\textit{Software}: \textsc{emcee} (\cite{emcee}), \textsc{batman} (\cite{batman}), \textsc{astropy} (\cite{astropy2018}), \textsc{PyAstronomy} (\cite{pya}), \textsc{matplotlib} (\cite{matplotlib}), \textsc{numpy} (\cite{numpy}), \textsc{pandas} (\cite{pandas}), \textsc{scipy} (\cite{scipy}), \textsc{astroARIADNE} (\cite{ariadne}) and \textsc{REBOUND} (\cite{rebound}), as well as their dependencies.

\section{Data Availability}
\label{sec:availability}
All RV measurements used in our joint model analysis have been included in this paper’s tables. The \tess{} data products from which we derived our light curves are publicly available online from the Mikulski Archive for Space Telescopes (MAST). Ground-based observations used in this paper is available from ExoFOP-TESS. CHEOPS data will be shared on reasonable request to the lead author.

\bibliographystyle{mnras}
\bibliography{ref}

\begin{table*}
\caption{Stellar Parameters for \target
   }
\label{tab:stellartable}
\renewcommand{\arraystretch}{1.25}
\begin{tabular}{l@{\hskip 0.5in}l@{\hskip 0.5in}l}
\toprule
Parameter & Value & Source
\\
\midrule

\addlinespace[4pt]
Catalog Information \\
\addlinespace[1pt]

~~~~R.A. (h:m:s)                      & \starRA     & Gaia DR3\\
~~~~Dec. (d:m:s)                      & \starDec    & Gaia DR3\\
~~~~Epoch							  & \starRefEpoch   & Gaia DR3 \\
~~~~Parallax (mas)                    & \starParallax  & Gaia DR3\\
~~~~$\mu_{ra}$ (mas yr$^{-1}$)        & \starPMRA   & Gaia DR3 \\
~~~~$\mu_{dec}$ (mas yr$^{-1}$)       & \starPMDec & Gaia DR3\\
~~~~Gaia DR3 ID                       & \starGaiaID   &  \\
~~~~TIC ID                            & \starTICID & \\
~~~~TOI ID                            & \starTOIID  & \\
\addlinespace[4pt]
Photometric properties \\
\addlinespace[1pt]
~~~~$TESS$ (mag)\dotfill            & \starTMag  & TIC v8.2         \\
~~~~$Gaia$ (mag)\dotfill            & \starGaiaMag & Gaia DR3               \\
~~~~Gaia RP (mag)\dotfill          & \starGaiaRPMag & Gaia DR3                 \\
~~~~Gaia BP (mag)\dotfill          & \starGaiaBPMag & Gaia DR3                 \\
~~~~$V_J$ (mag)\dotfill             & \starVMag & APASS DR10      \\
~~~~$B_J$ (mag)\dotfill             & \starBMag & APASS DR10      \\
~~~~$J$ (mag)\dotfill               & \starJMag & 2MASS           \\
~~~~$H$ (mag)\dotfill               & \starHMag & 2MASS           \\
~~~~$K_s$ (mag)\dotfill             & \starKMag & 2MASS           \\
~~~~$W_1$ (mag)\dotfill             & \starwoneMag & WISE           \\
~~~~$W_2$ (mag)\dotfill             & \starwtwoMag & WISE           \\
\addlinespace[4pt]
Derived Properties \\
\addlinespace[1pt]

~~~~\mstar (\msun)\dotfill      &  \starMass & Constant Period Circular Fit\\
~~~~\rstar (\rsun)\dotfill      & \starRadius  & Constant Period Circular Fit      \\
~~~~$\loggstar$ (cgs)\dotfill       & \starLogg  & Constant Period Circular Fit      \\
~~~~$\lstar$ ($\lsun$)\dotfill      & \starLuminosity  & Constant Period Circular Fit    \\
~~~~$\teffstar$ (K)\dotfill        &  \starTeff  & SED\\
~~~~$\feh$ (dex)\dotfill            & \starfeh & SED\\
~~~~Distance (pc)\dotfill           & \starDistance & SED\\
~~~~\rhostar (\gcmc)\dotfill &  \starRho  & Constant Period Circular Fit\\
~~~~Age (Gyr) \dotfill &  \starAge  & SED\\
\addlinespace[4pt]
Limb-darkening coefficients \\
\addlinespace[1pt]
~~~$u_{\mathrm{1,TESS}}$               \dotfill    & \starTESSuOne     &  \\
~~~$u_{\mathrm{2,TESS}}$               \dotfill    &  \starTESSuTwo    & \\
~~~$u_{\mathrm{1,z_s}}$               \dotfill    & \starzsuOne     & \\
~~~$u_{\mathrm{2,z_s}}$               \dotfill    &  \starzsuTwo     & \\  
~~~$u_{\mathrm{2,CHEOPS}}$           \dotfill    &  \starcheopsuTwo     & \\  
~~~$u_{\mathrm{2,CHEOPS}}$           \dotfill    &  \starcheopsuTwo     & \\  
\addlinespace[4pt]
Scaling coefficients \\
\addlinespace[1pt]
~~~$c_{\mathrm{TESS}}$               \dotfill    & \cTess     &  \\
~~~$c_{\mathrm{McD}}$               \dotfill    & \cMcd     &  \\
~~~$c_{\mathrm{Teid}}$               \dotfill    & \cTeid     &  \\
~~~$c_{\mathrm{CTIO, 1}}$               \dotfill    & \cCtioOne     &  \\
~~~$c_{\mathrm{CTIO, 2}}$               \dotfill    & \cCtioTwo     &  \\
~~~$c_{\mathrm{CHEOPS, 1}}$               \dotfill    & \cCheopsOne     &  \\
~~~$c_{\mathrm{CHEOPS, 2}}$               \dotfill    & \cCheopsTwo     &  \\
~~~$c_{\mathrm{CHEOPS, 3}}$               \dotfill    & \cCheopsThree     &  \\
\bottomrule
\addlinespace[4pt]
\end{tabular}

\raggedright

\end{table*}

\begin{table*}
\caption{Planet Parameters for \target}
\label{tab:planettable}
\renewcommand{\arraystretch}{1.5}
\begin{tabular}{l@{\hskip 0.15in}l@{\hskip 0.15in}l@{\hskip 0.15in}l@{\hskip 0.15in}l}
\toprule
Parameter & \multicolumn{4}{c}{Constant Period}\\
& \multicolumn{2}{c}{Circular} & \multicolumn{2}{c}{Eccentric}  \\
& \multicolumn{1}{c}{\bPlanet} & \multicolumn{1}{c}{\cPlanet} & \multicolumn{1}{c}{\bPlanet} & \multicolumn{1}{c}{\cPlanet}  \\
\midrule
Light curve parameters \\
$P$ (days) \dotfill & \bPeriod & \cPeriod & \bPeriodecc & \cPeriodecc \\
$T_c$ (${\rm BJD} - 2457000$) \dotfill & \bEpoch & \cEpoch & \bEpochecc & \cEpochecc \\
$T_{14}$ (hr) \dotfill & \bDuration & \cDuration & \bDurationecc & \cDurationecc  \\
$T_{12} = T_{34}$ (min) \dotfill & \bIngressDuration & \cIngressDuration & \bIngressDurationecc & \cIngressDurationecc \\
$\arstar$ \dotfill & \bAOR & \cAOR & \bAORecc & \cAORecc \\
$\rpl/\rstar$ \dotfill & \bROR & \cROR  & \bRORecc & \cRORecc  \\
$b \equiv a \cos i/\rstar$ \dotfill & \bImpactParameter & \cImpactParameter  & \bImpactParameterecc & \cImpactParameterecc  \\
$i$ (deg) \dotfill & \bInclination & \cInclination  & \bInclinationecc & \cInclinationecc \\
$e$ \dotfill & 0 (Fixed) & 0 (Fixed) & \bEccentricityecc & \cEccentricityecc\\
$\omega$ (deg) \dotfill & 0 (Fixed) & 0 (Fixed) & \bOmegaecc & \cOmegaecc\\
\midrule
Planetary parameters \\
$\rpl$ ($\rearth$) \dotfill & \bRadius & \cRadius  & \bRadiusecc & \cRadiusecc \\
$a$ (AU) \dotfill & \bSemimajorAxis & \cSemimajorAxis & \bSemimajorAxisecc & \cSemimajorAxisecc \\
$T_{\rm eq}$ (K) \dotfill & \bTeq & \cTeq & \bTeqecc & \cTeqecc \\
$\langle F \rangle$ ($S_{\earth}$) \dotfill & \bIrradiation & \cIrradiation & \bIrradiationecc & \cIrradiationecc  \\
\bottomrule
\end{tabular}%

\end{table*}

\clearpage
\newpage

{ \itshape \footnotesize \noindent
$^{1}$ University of Southern Queensland, West St, Darling Heights, Toowoomba, Queensland, 4350, Australia \\
$^{2}$Department of Physics and Kavli Institute for Astrophysics and Space Research, Massachusetts Institute of Technology, 77 Massachusetts Ave, Cambridge, MA 02139, USA \\
$^{3}$ Sternberg Astronomical Institute Lomonosov Moscow State University 119992, Moscow, Russia, Universitetskii prospekt, 13 \\
$^{4}$ Center for Astrophysics \textbar \ Harvard \& Smithsonian, 60 Garden Street, Cambridge, MA 02138, USA \\
$^{5}$ NASA Exoplanet Science Institute - Caltech/IPAC, 1200 E. California Blvd, Pasadena, CA 91125 USA \\
$^{6}$ George Mason University, 4400 University Drive, Fairfax, VA, 22030 USA \\
$^{7}$ NSF's National Optical-Infrared Astronomy Research Laboratory, 950 N. Cherry Avenue, Tucson, AZ 85719, USA \\
$^{8}$ NASA Ames Research Center, Moffett Field, CA 94035, USA \\
$^{9}$ Instituto de Astrof\'isica de Canarias (IAC), E-38205 La Laguna, Tenerife, Spain \\
$^{10}$ Departamento de Astrof\'isica, Universidad de La Laguna (ULL), E-38206 La Laguna, Tenerife, Spain \\
$^{12}$ Kotizarovci Observatory, Sarsoni 90, 51216 Viskovo, Croatia \\
$^{13}$ Department of Earth, Atmospheric, and Planetary Sciences, Massachusetts Institute of Technology, Cambridge, MA 02139, USA \\
$^{14}$ Department of Aeronautics and Astronautics, Massachusetts Institute of Technology, Cambridge, MA 02139, USA \\
$^{15}$ Department of Astrophysical Sciences, Princeton University, Princeton, NJ 08544, USA \\
$^{16}$ Department of Astronomy, MC 249-17, California Institute of Technology, Pasadena, CA 91125, USA \\
$^{17}$ Department of Physics and Astronomy, University of New Mexico, 210 Yale Blvd NE, Albuquerque, NM 87106, USA \\
$^{18}$ Center for Computational Astrophysics, Flatiron Institute, 162 5th Ave New York, NY 10010, USA \\
}

\bsp	
\label{lastpage}

\end{document}